\documentclass[a4paper,11pt]{article}
\pdfoutput=1 
\usepackage{jcappub}
\usepackage{graphicx}
\usepackage{color}
\usepackage{makecell}
\usepackage{afterpage}
\usepackage[dvipsnames]{xcolor}
\usepackage{multirow}
\usepackage[noabbrev]{cleveref}
\usepackage[normalem]{ulem}		
\usepackage{xspace}
\usepackage{subcaption}
\usepackage{booktabs}
\usepackage{makecell}

\usepackage{centernot}
\usepackage{mathtools}
\usepackage{stmaryrd}
\usepackage[utf8]{inputenc}

\usepackage{csquotes} 

\usepackage{placeins}
\usepackage{titlesec}
\titlespacing*{\section} {0pt}{1.5ex plus 0.5ex minus .2ex}{-1.8ex plus .2ex}
\titlespacing*{\subsection} {0pt}{0.75ex plus 0.5ex minus .2ex}{-1.8ex plus .2ex}

\newcommand{\dd}{\mathrm{d}}
\newcommand{\fulleqref}[1]{equation~\eqref{#1}}

\newcommand{\fullmanyeqref}[1]{equations~\eqref{#1}}

\newcommand{\tquote}[1]{``#1''}

\newcommand{\abs}[1]{\ensuremath{\lvert#1\rvert}}

\newcommand{\parderconst}[3]{\ensuremath{\left(\frac{\partial #1}{\partial #2}}\right)_#3}
\newcommand{\der}[2]{\ensuremath{\frac{\mathrm{d} #1}{\mathrm{d} #2}}}

\newcolumntype{L}[1]{>{\raggedright\arraybackslash}p{#1}}
\newcolumntype{C}[1]{>{\centering\arraybackslash}p{#1}}

\newcommand{\class}{{\sc class}\xspace}

\newcommand{\Neff}{\ensuremath{N_\mathrm{eff}}\xspace}
\newcommand{\lcdm}{\ensuremath{\Lambda\mathrm{CDM}}\xspace}
\newcommand{\can}{\mathrm{CanDM}}
\newcommand{\sigmavsq}{\ensuremath{\langle\sigma_{32} v^2\rangle}}
\newcommand{\DNeff}{\ensuremath{\Delta N_\mathrm{eff}}\xspace}
\newcommand{\nnr}{\ensuremath{n_\mathrm{NR}}\xspace}
\newcommand{\rnr}{\ensuremath{\rho_\mathrm{NR}}\xspace}
\newcommand{\pnr}{\ensuremath{P_\mathrm{NR}}\xspace}

\newcommand{\alphcandm}{\ensuremath{\alpha}}
\newcommand{\mcandm}{\ensuremath{m}}
\newcommand{\fcandm}{\ensuremath{f}}
\newcommand{\lnalph}{\ensuremath{\log_{10} \alphcandm}}
\newcommand{\lnM}{\ensuremath{\log_{10} \mcandm}}

\title{Cannibalism hinders growth: Cannibal Dark Matter and the $\mathbf{S_8}$ tension}

\author[1]{Stefan Heimersheim,} 
\author[2]{Nils Sch\"oneberg,}
\author[3]{Deanna C. Hooper,}
\author[2]{Julien Lesgourgues}

\affiliation[1]{Institute of Astronomy, \\ University of Cambridge, Madingley Road, Cambridge CB3 0HA, UK.}
\affiliation[2]{Institute for Theoretical Particle Physics and Cosmology (TTK), \\ RWTH Aachen University, D-52056 Aachen, Germany.}
\affiliation[3]{Service de Physique Th\'eorique, CP225, \\ Universit\'e Libre de Bruxelles, Boulevard du Triomphe, 1050 Bruxelles, Belgium.}

\emailAdd{heimersheim@ast.cam.ac.uk}
\emailAdd{schoeneberg@physik.rwth-aachen.de}

\abstract{Many models of dark matter have been proposed in attempt to ease the $S_8$ tension between weak lensing and CMB experiments. One such exciting possibility is cannibalistic dark matter (CanDM), which has exothermal number-changing interactions allowing it to stay warm far into its non-relativistic regime.
Here we investigate the cosmological implications of CanDM and how it impacts CMB anisotropies and the matter power spectrum, by implementing the model within a linear Einstein-Boltzmann solver. We show that CanDM suppresses the small scale matter power spectrum in a way very similar to light Warm Dark Matter or Hot Dark Matter. However, unlike in those models, the suppression may happen while the CanDM model still remains compatible with CMB constraints. We put strong constraints on the interaction strength of CanDM as a function of its abundance for both constant and temperature-dependent thermally-averaged cross sections. We find that the CanDM model can easily solve the $S_8$ tension (but has no impact on the Hubble tension). Indeed, it can accommodate values of $S_8$ of the order of 0.76 while being compatible with CMB+BAO data. However, as long as the $S_8$ tension remains moderate, the overall $\chi^2$ improvement is relatively small given the number of extra free parameters, and the CanDM model is not significantly preferred. 
}

\begin{document}

\hfill{\small TTK-20-26}

\hfill{\small ULB-TH/20-11}

\vspace*{-2\baselineskip}

\maketitle

\setlength{\parskip}{\baselineskip}%
\section{Introduction}
The current standard cosmological model (\lcdm) describes a universe largely dominated by dark matter and dark energy, and is supported by a wealth of evidence across many different cosmological scales. Observations ranging from Big Bang Nucleosynthesis (BBN)~\cite{Aver:2015iza, Peimbert:2016bdg, Izotov:2014fga,Cooke:2017cwo} and Cosmic Microwave Background (CMB)~\cite{Planck2018} to late time probes such as Baryon Acoustic Oscillations (BAO)~\cite{Beutler_2011, Ross:2014qpa, Alam:2016hwk}, luminosities of supernovae of type Ia (SNIa)~\cite{Scolnic:2017caz}, and the clustering of small scale structure observed in galaxy catalogues and weak lensing experiments~\cite{Heymans:2013fya, Abbott:2020knk, Joudaki:2019pmv} all point towards the existence of a collisionless non-relativistic particle species, playing the role of cold dark matter (CDM).

Despite its many predictions and successes, \lcdm cosmology still exhibits tensions across different datasets; the most notable is the so-called $H_0$ tension~\cite{Freedman:2017yms, Aylor:2018drw}. It appears when comparing local measurements of the expansion rate of the universe, such as those obtained by the SH0ES collaboration \cite{Riess+2019}, with the value calculated by CMB experiments~\cite{Planck2018}. Another tension, albeit slightly milder, is the discrepancy that appears~\cite{Chang:2018rxd,MacCrann:2014wfa} when measuring the clustering of small scale structure, often quantified with the $S_8$ parameter: in this case, weak lensing experiments such as DES~\cite{Abbott:2020knk,Zuntz:2017pso,Drlica-Wagner:2017tkk}, KiDS~\cite{Hildebrandt:2018yau, Wright:2018nix, Joudaki:2019pmv, Asgari:2019fkq,Asgari:2020wuj,Heymans:2020gsg}, and CFHTLens~\cite{Heymans:2013fya} all find lower values of $S_8$ than those calculated by CMB experiments assuming \lcdm cosmology. A recent analysis of \cite{Joudaki:2019pmv} has obtained $S_8=0.762^{+0.025}_{-0.024}$, which is in $2.3\sigma$ tension with the Planck 2018~\cite{Planck2018} value of $S_8=0.825\pm0.011$. 
We will adopt this measurement as our baseline $S_8$ data. We note, however, that by using slightly different assumptions and data, reference \cite{Asgari:2019fkq} gets a stronger (3.2$\sigma$) tension; while \cite{Heymans:2020gsg} also gets a $\sim 3 \sigma$ tension when combining KiDS-1000 weak lensing data with BOSS galaxy clustering data.
These tensions have motivated the exploration of dark matter models beyond the standard CDM paradigm.

One such class of interesting beyond-CDM models are those in which the dark matter undergoes \enquote{cannibalistic} number changing $3\rightarrow2$ interactions. These models, known as Cannibal Dark Matter (CanDM), were initially proposed by \cite{Dolgov:1980abc,Carlson+1992,Dolgov:2017ujf} in the context of self-interacting dark matter \cite{Machacek1994, deLaix+1995, Dolgov:1995rm}.
They have seen a resurgence of interest in the last few years~\cite{Hochberg:2014kqa, Kuflik+2016, Pappadopulo+2016, Farina+2016,Erickcek:2020wzd} given their potential to not only mitigate the small-scale crisis~\cite{Spergel:1999mh, deBlok:2009sp, Boylan_Kolchin_2011,Salucci:2018hqu} but also to alleviate the $S_8$ tension~\cite{BuenAbad+2018}.
Additionally, these models present a rich phenomenology, occurring naturally in many BSM scenarios~\cite{Bernal:2015ova,Chu:2017msm,Bernal:2018hjm,Heeba:2018wtf}.

To study the impact of CanDM on different cosmological observables, we have performed the first full implementation of this model in an Einstein-Boltzmann solver. We use this to analyse the effects this model has on the CMB and large-scale structure formation, with a special interest on its potential to alleviate the $S_8$ tension. We will further study the important role the mass of the CanDM particle plays in modeling this behavior, leading to several regimes with different phenomenological impact.

This paper is organized as follows: in section \ref{sec:method} we introduce the model and discuss the background and perturbation equations in the CanDM model.
In section \ref{sec:implications} we describe the implications of this model for CMB power spectra and the matter power spectrum suppression.
In section~\ref{sec:results}, we show the parameter constraints from cosmological probes and the effect of CanDM on the $S_8$ tension.
Finally, in section~\ref{sec:conclusion} we summarize our findings and discuss the future implications of this work.

\section{A cosmological model with Cannibal Dark Matter}
\label{sec:method}

\subsection{Cannibal species}
\label{subsec:model}

Before we discuss the details of this model we want to stress
that the formalism discussed below does not depend on the precise form of the Lagrangian. Any model that allows for a number changing interaction will yield similar results. Within this work we only consider models featuring $3\leftrightarrow2$ interactions and following a Bose-Einstein distribution in local thermodynamic equilibrium. Nonetheless, we have checked that fermionic models lead to small ($\sim 10\%$) changes in the relativistic regime, but do not significantly alter the overall thermal evolution.

To exemplify our phenomenological analysis of CanDM we use the toy model proposed in \cite{BuenAbad+2018}. This model just contains a single decoupled scalar field $\phi$ with mass~$m$, giving rise to the Lagrangian 
\begin{gather}
\mathcal{L_\mathrm{CanDM}}= \frac12 \left(\partial^\mu\phi\right) \left(\partial_\mu\phi\right)
- \frac12 m^2 \phi^2
- m \kappa \frac{\phi^3}{3!}
- \lambda^2 \frac{\phi^4}{4!}~.
\label{eq:lagrangian}
\end{gather}
This Lagrangian allows for $2\leftrightarrow2$ interactions ($\phi \phi \leftrightarrow \phi \phi \sim \lambda^2 \mathrm{\,or\,} \kappa^2$) potentially ensuring thermal equilibrium, and $3\leftrightarrow2$ interactions ($\phi \phi \phi \leftrightarrow \phi \phi \sim \kappa \lambda^2 \mathrm{\,or\,} \kappa^3$) potentially establishing chemical equilibrium\footnote{Any $1\leftrightarrow N$ process is kinematically forbidden for nonzero mass, while higher order interactions are disfavored by additional factors of $\kappa\mathrm{\,or\,}\lambda$, which are assumed to be small in this toy model.}.
We will assume that CanDM has been produced by a mechanism such as freeze-in, freeze-out, or during re-heating after an inflationary stage. The precise nature of the production mechanism does not matter for the purpose of this paper, provided it occurred before the era relevant for CMB physics and large scale structure formation, and provided that CanDM is decoupled from other species throughout this era.

Similarly to \cite{BuenAbad+2018,Erickcek:2020wzd} we parametrize the thermally averaged cross section in the non-relativistic 
regime with a dimensionless interaction strength parameter $\alpha$, stressing the 
independence from the specific form of the Lagrangian. Furthermore, we allow for a scaling of the cross section with the CanDM temperature $T$ of the form\footnote{We formulate the dependence with respect to some reference scale $T_*$ (such as the mass, the initial temperature or just some units). In our implementation we use the initial CanDM temperature at $a=10^{-14}$ as the reference scale.} 
$\sigmavsq\propto T^n$:
\begin{align}
	\sigmavsq \approx \frac{\alpha^3}{m^5} \left(\frac{T}{T_*}\right)^n &\implies \Gamma_{3\to 2} \approx \frac{\alpha^3}{m^5} n_\mathrm{NR}^2 \left(\frac{T}{T_*}\right)^n~.
	\label{eq:gamma32}
\end{align}
This parameterization applies to a variety of models beyond the particular case of the Lagrangian presented in equation \eqref{eq:lagrangian}\footnote{The Lagrangian in \fulleqref{eq:lagrangian} with $\kappa=\lambda$ corresponds to $\alpha\sim \lambda^2/(4\pi)$ and $n=0$ \cite{BuenAbad+2018}. Note that the preferred parameter region of $\alpha\gg1$ ($\lambda\gg 1$) is incompatible with the simple toy model; a UV completion is discussed in \cite{BuenAbad+2018}.}. In particular, we consider two temperature dependencies motivated by different cannibalistic models. In the first model, motivated by \cite{Hochberg:2015vrg} (eq. 7.1) and \cite{BuenAbad+2018} (eq. 14), we treat the cross section as constant ($n=0$); while in the second model, motivated by \cite{Hochberg:2014kqa} (eq. 15), we investigate a cross section proportional to $T^2$ ($n=2$).

We will consider cases in which CanDM plays the role of Dark Matter, or coexists with a plain decoupled CDM species. In both cases, the impact of CanDM on the cosmological evolution depends on four independent quantities: the cannibalistic fraction $f$ of Dark Matter,  with $0\leq f \leq1$; the CanDM mass $m$; and the parameters  $\alpha$ and $n$ describing the rate $\Gamma_{3\to2}$\,, which only affect the non-relativistic phase of the evolution, as we shall see in the next section. One could expect the initial CanDM temperature $T_{\rm ini}$ to be a fifth parameter; however $T_{\rm ini}$ is not independent from $(m, f, \alpha, n)$ and can be inferred from the other parameters. 

\subsection{Background evolution}
\label{subsec:bgevo}
As long as the species is ultra-relativistic, the $3\rightarrow2$ process is well balanced by the $2\rightarrow3$ process, since the gap between the initial and final states is negligible compared to the total energy involved in the reactions. This establishes chemical equilibrium and conservation of the comoving number density. Eventually, when the momentum of individual particles becomes of the order of their mass, the energy difference between final and initial states becomes relevant, and the $3\rightarrow2$ process remains efficient while the $2\rightarrow3$ process freezes out. After this point, the comoving number density of CanDM is no longer conserved. The $3\rightarrow2$ process heats the species as the mass of the disappearing particles gets converted into additional kinetic energy for the remaining particles; this inspired the name of \enquote{cannibal} dark matter. Since the interaction rate is proportional to the squared number density, number changing processes will eventually freeze out. When this happens, chemical equilibrium is lost and the chemical potential becomes non-negligible.

On the other hand, we assume that the $2\rightarrow2$ process is very efficient and enforces thermal and kinetic equilibrium at any time, such that the unperturbed phase-space distribution can be described by a Bose-Einstein or Fermi-Dirac distribution. Thus, at the homogeneous level, CanDM can be completely described by its chemical potential $\mu$ and temperature $T$. In our bosonic toy model we can write the unperturbed phase-space distribution at any time as
\begin{equation}
f(p,\mu,T) = \frac{1}{\exp\left((\sqrt{p^2+m^2}-\mu)/T\right)-1}~.
\label{eq:psd}
\end{equation}
As such, we will always need two background equations to describe the homogeneous evolution of the CanDM species: one for $\mu$ and one for $T$.

\begin{figure}[t]
	\centering
	\includegraphics[scale=0.75]{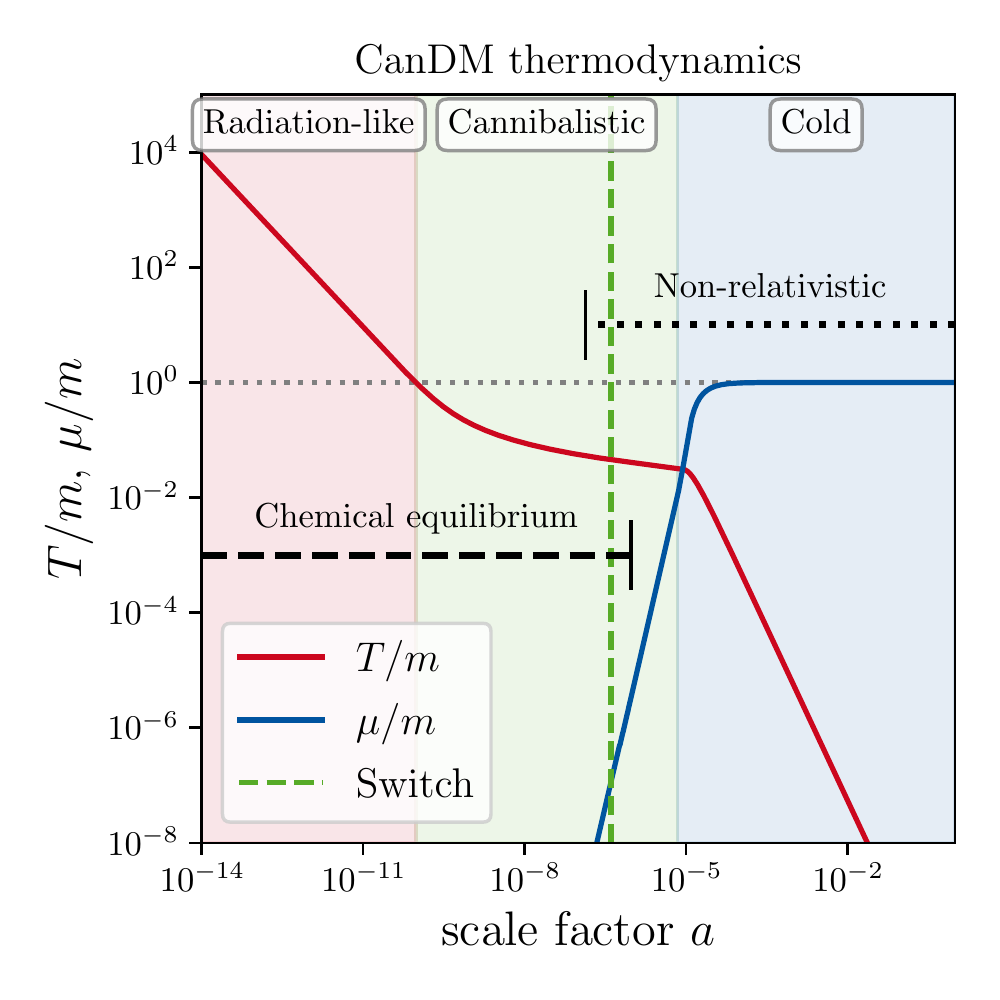}
	\includegraphics[scale=0.75]{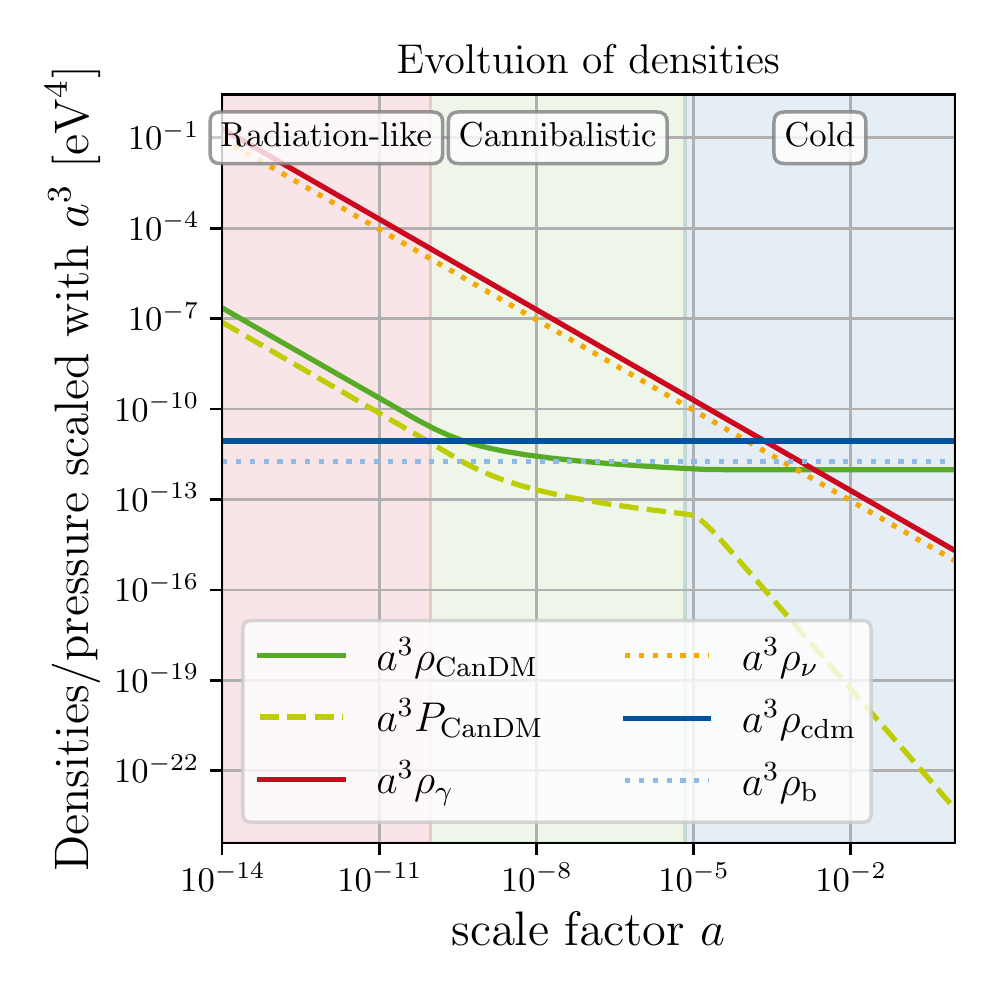}
	\caption{CanDM evolution for $m=100\,\mathrm{keV}$, $\alpha=10, n=0$, and $f=0.1$.
		\textbf{Left:} Temperature $T$ and chemical potential $\mu$ of CanDM as a function of the scale factor~$a$, illustrating the three different regimes of CanDM evolution: \textit{radiation-like}, \textit{cannibalistic}, and \textit{cold}.
		The chemical equilibrium (CE) and non-relativistic (NR) regions are shown as
		black dashed/dotted bars. The green dashed line shows the time at which the code switches between the CE and NR approximations.
	\textbf{Right:} Typical evolution of densities of CanDM, CDM, baryons, neutrinos, radiation, and CanDM pressure (all multiplied by $a^3$). }
	\label{fig:T_rho_background}
\end{figure}

\noindent
Figure \ref{fig:T_rho_background} illustrates the typical background evolution and thermal history of CanDM:

\begin{itemize}
\item At early times and high temperatures, the mass is negligible compared to the average momentum of CanDM particles, which behave as ordinary {\it radiation}: $T\propto a^{-1}$ and $\rho\propto a^{-4}$. Additionally the $2\to3$ and $3\to2$ processes are equally efficient and maintain chemical equilibrium (CE) with $\mu\approx0$. 
\item When the temperature reaches $T \sim m$, the $2\to3$ process freezes out and the \textit{cannibalistic} phase begins. We will see that in this regime, the self-heating makes the temperature decrease only logarithmically, $T\sim 1/ \log a$, while the density drops as $\rho \propto a^{-3}/\log a$. Thus the cannibalistic phase features a very gradual transition of CanDM particles from the relativistic to non-relativistic regime, with $T/m$ decreasing like $(\log a)^{-1}$ instead of the usual $a^{-1}$.
\item As the universe expands and the number density of CanDM decreases, also the ratio $\Gamma_{3\to 2}/H$ decreases. Finally, when $\Gamma_{3\to2}<H$, the species behaves like an ordinary \textit{cold} species with $T\propto a^{-2}$, $\rho\propto a^{-3}$, and $\mu \to m$\footnote{This result can be analytically derived from entropy and number count conservation for a decoupled non-relativistic species, since $(\mu-m)/T$ needs to be constant for such a species, and $T \propto a^{-2}$.}.
\end{itemize}

The thermal history and background evolution of particles with number changing interactions is not straightforward to derive from the Boltzmann equation\footnote{For a detailed and instructive derivation of the moments of the Boltzmann equation, we refer the reader to Appendix A of \cite{BuenAbad+2018}.}. Fortunately, it was shown in \cite{BuenAbad+2018} that even if exact equations of evolution are complicated and unpractical, a set of simplified equations can be derived in two regimes: the chemical equilibrium (CE) epoch, and the non-relativistic (NR) epoch. For the typical models in which we are interested, these two epochs have a significant overlap, as can be seen in the left panel of figure \ref{fig:T_rho_background}. Our strategy is then to swap from one approximation to the other at some intermediate time marked with a dashed vertical line in figure \ref{fig:T_rho_background}. Models without an overlap are those in which the rate $\Gamma_{3\to2}$ is so small that $3\to2$ interactions freeze out already in the relativistic regime. In that case, there is no cannibalistic phase -- or at most a very short one. DM behaves like a simple massive fluid, with a quick transition from the radiation-like to the cold phase, and no -- or very little -- impact of number-changing interactions. Since we are interested in the cosmological signature of the cannibalistic behaviour, we will exclude this case from our analysis.

Within the CE phase, the efficiency of the $3\rightarrow2$ process implies that the chemical potential of CanDM is negligible. We can then work in the approximation $\mu=0$ in which, for a known temperature $T$, we obtain the energy density $\rho(T)$ and pressure $P(T)$ by integrating the phase-space distribution (\ref{eq:psd}) over momentum. To get an equation of evolution for $T$, we start from the second moment of the Boltzmann equation (see appendix \ref{ap:conservationLTE}), which gives the usual energy conservation law that applies to any species not interacting with other ones:
\begin{equation}
\der{\rho}{\log a} + 3 (\rho+P) = 0~.
\label{eq:ec}
\end{equation}
Indeed, the term accounting for number changing processes only appears explicitly in the first moment of the Boltzmann equation, that is, in the evolution equation of the number density. Equation~(\ref{eq:ec}) can be written as an equation for the evolution of temperature:
\begin{equation}
	\der{\ln T}{\ln a} = \frac{-3(\rho(T)+P(T))}{\der{\rho}{\ln T}(T)}~.
	\label{eq:Tev}
\end{equation}
Like $\rho(T)$ and $P(T)$, the term $\der{\rho}{\ln T}(T)$ can be explicitly obtained from a numerical integral over momentum. It is thus possible to solve the differential equation (\ref{eq:Tev}) and get the full evolution of $T(a)$, $\rho(a)$ and $P(a)$ starting from an initial temperature $T_{\rm ini}$\,. This temperature is in principle a free parameter of the model, but in our numerical implementation we use a shooting method to eliminate it in favour of the final cannibalistic fraction $f$ for each choice of $(m, \alpha, n)$.

As soon as CanDM becomes non-relativistic (NR),  the phase-space distribution can be approximated as a Maxwell-Boltzmann distribution:
\begin{equation}
	f_\mathrm{NR}(p,\mu,T) \approx \frac{1}{\exp \left((\sqrt{p^2+m^2}-\mu)/T\right)} = \exp (\mu/T) \exp (-\sqrt{p^2+m^2}/T)~.
\end{equation}
Mathematically, the approximation is in $E-\mu \gg T$, but in appendix \ref{ap:approximationvalidity} we confirm the validity of this approximation for $T \ll m$. Thus it can be used both within the CE phase (when $\zeta \equiv \mu/T\ll1$) and afterwards (when $\zeta$ becomes sizeable). In this limit, the expressions for number density, energy density and pressure can be derived analytically as a function of $\zeta$ and $x = m/T$ (see sec. 9.6.23 of \cite{Abramowitz+1964}):
\begin{equation}
\begin{alignedat}{2}
\nnr&=m^3 \frac{e^\zeta}{2\pi^2} \frac{K_2(x)}{x}~,\\
\rnr&=m^4 \frac{e^\zeta}{2\pi^2} \left(\frac{3 K_2(x)}{x^2} + \frac{K_1(x)}{x}\right)~,\\
\pnr&=m^4 \frac{e^\zeta}{2\pi^2} \frac{K_2(x)}{x^2} = T \nnr~,
\label{eq:bessel_approximation}
\end{alignedat}
\end{equation}
where $K_i(x)$ is the $i$-th modified Bessel function of the second kind. The evolution of $\zeta$ and $x$ can be inferred from the first two moments of the Boltzmann equation, which are  the number density evolution equation (see appendix \ref{ap:conservationLTE})
\begin{equation}
\der{n}{\log a} +  3n =  \dot{N}[f]/H
\label{eq:nc}
\end{equation}
and the energy conservation equation (\ref{eq:ec}).
The number density current $\dot{N}[f]$ is induced by $3\rightarrow2$ interactions. Its general expression (appendix \ref{ap:conservationLTE}, equation \eqref{eq:ap:dotN}) is very challenging to evaluate, as it depends on an integral over the product of multiple phase-space functions, which would require a computationally expensive numerical integration. Fortunately, in the non-relativistic regime, it has a trivial limit
\begin{equation}
	\dot{N}[f]_\mathrm{NR} = - \sigmavsq \, \nnr^2 \, \cdot \left(e^{\mu/T}-1\right) \equiv \left(1-e^{\mu/T}\right) \Gamma_{3\to 2}~,
\end{equation}
where $\Gamma_{3\to 2}$ is taken from (\ref{eq:gamma32}). After plugging the non-relativistic quantities (\ref{eq:bessel_approximation}) in the conservation equations (\ref{eq:ec}, \ref{eq:nc}), one can derive a system of evolution equations for $\zeta$ and $x$, 
\begin{equation}
\begin{alignedat}{1}
\der{\zeta}{\ln a}&=x^2\der{\ln \left(\rnr e^{-\zeta} \right)}{x} \der{x}{\ln a} - 3 (1+\pnr/\rnr)~,\\
\der{x}{\ln a}&=\frac{\Gamma_{3\to 2}/H-3 \pnr/\rnr}{-x^2 \dd\ln (\rnr/m\nnr)/\dd x}~.
\end{alignedat}
\label{eq:zx}
\end{equation}
Given the explicit analytical expressions
\begin{equation}
\begin{alignedat}{1}
\der{\ln \left(\rnr e^{-\zeta}\right)}{x} &= \frac{(x^3+12x) K_0(x) + (5x^2+24)K_1(x)}{3K_2(x)+x K_1(x)}~,\\
\der{\ln (\rnr/m\nnr)}{x} &= 3- \frac{x^2}{2} \frac{K_1(x)^2 + K_1(x)K_3(x)-K_0(x)K_2(x) - K_2(x)^2}{K_2(x)^2}~,
\end{alignedat}
\end{equation}
\enlargethispage{2\baselineskip}
we can solve the system (\ref{eq:zx}) to infer $\zeta(a)$ and $x(a)$, and thus the full thermal history and background evolution of CanDM.

We implemented these equations and the automatic switch from the CE to NR approximation in the background module of the Boltzmann solver \textsc{class} \cite{Blas_2011}.
Note that the differential system (\ref{eq:zx}) is stiff, especially deep in the NR regime, when $\mu \to m$. We solve it with an implicit integration method, namely, the integrator \texttt{ndf15} introduced in \cite{Blas_2011} to deal with the perturbation equations of \textsc{class}. For the purposes of this work, we propagated this method to the background equations. In \textsc{class v3.0}~\cite{Lucca:2019rxf}, the \texttt{ndf15} integrator will become available by default not only for perturbation equations, but also for background and thermodynamical equations.

\begin{figure}[t]
	\centering
	\includegraphics[scale=0.75]{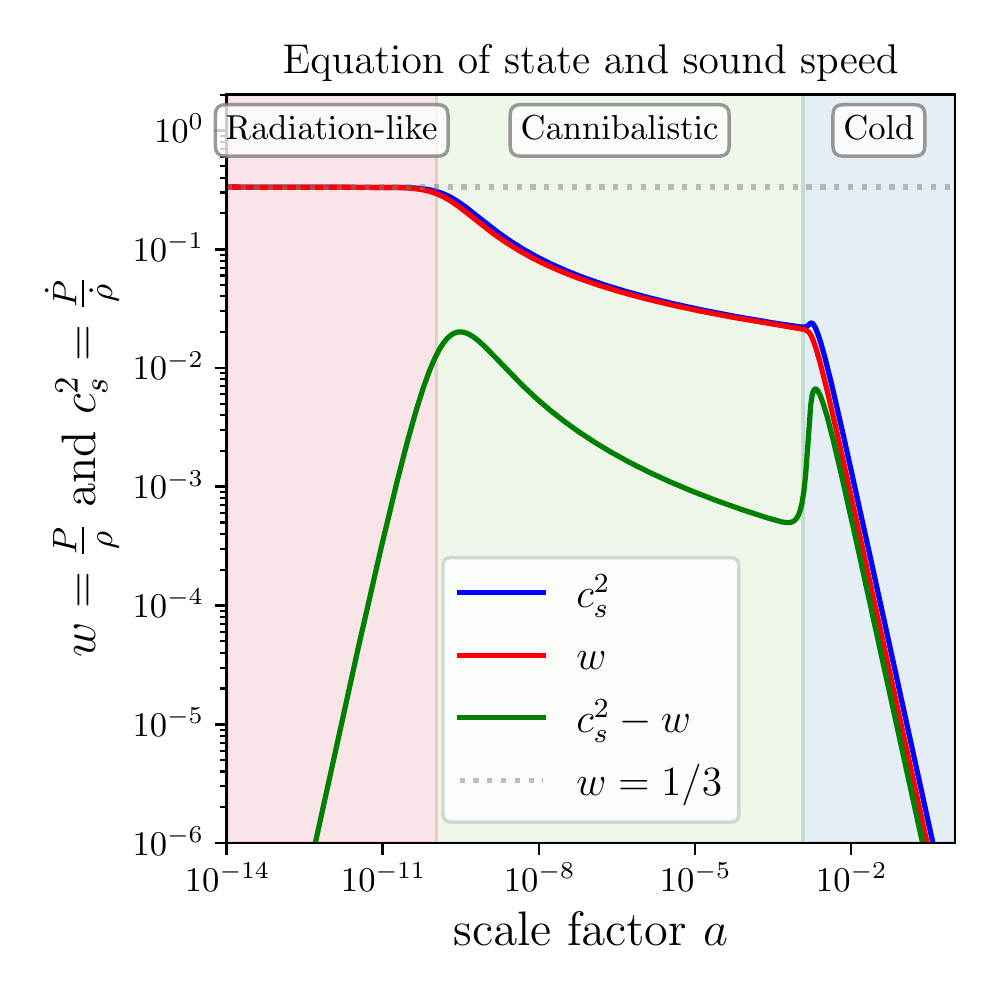}
	\includegraphics[scale=0.75]{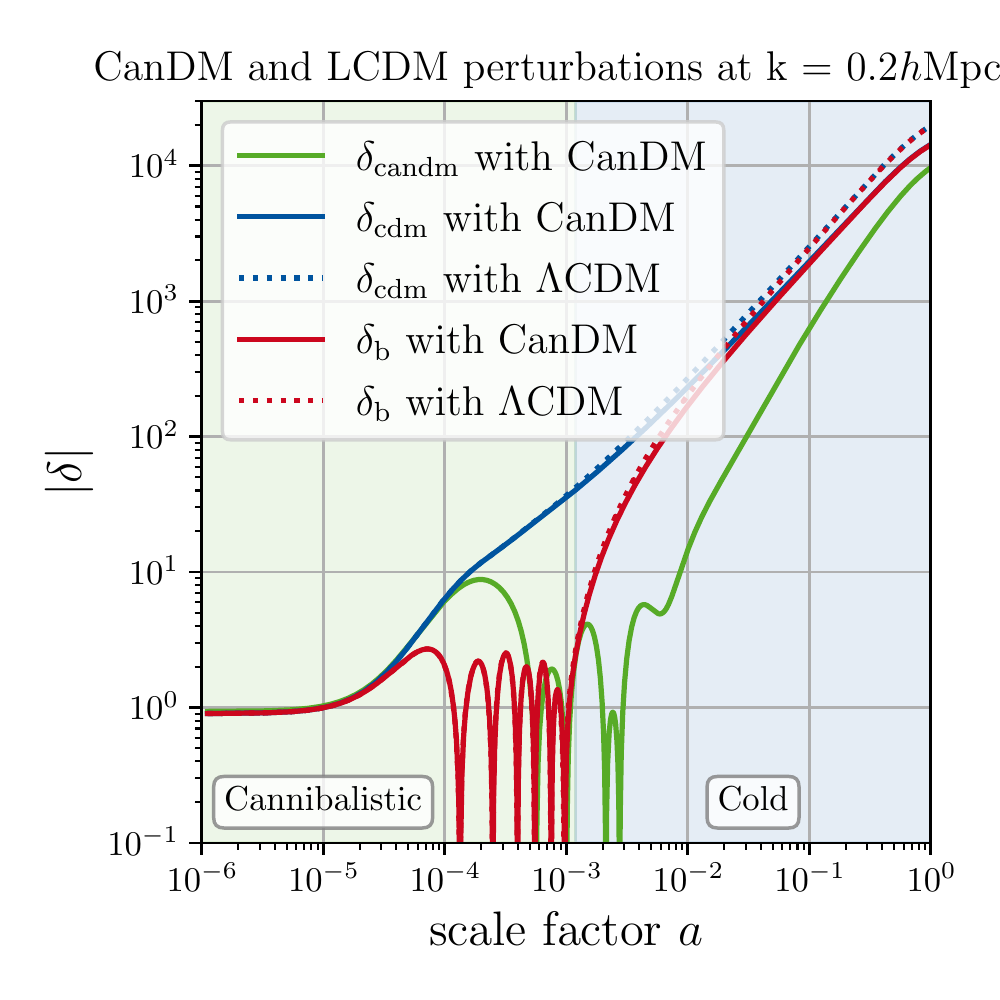}
	\caption{Effects of $f=10\%$ CanDM with $m=100\,\mathrm{keV}, \alpha=10^4$ and $n=0$.
		\textbf{Left:} The equation of state $w=P/\rho$ and sound speed $c_s^2=P'/\rho'$ as a function of scale factor. \textbf{Right:} The matter perturbations for CDM, baryons, and CanDM as a function of scale factor.
		The solid lines correspond to CanDM, while the dotted lines show the $\Lambda$CDM model.
	}
	\label{fig:perturbs_eos}
\end{figure}
\subsection{Perturbation evolution}
\label{subsec:ptevo}
With these approaches, we have derived the background evolution of the CanDM species. The further derivation of the perturbation equations is then very simple and performed in appendix~A of \cite{BuenAbad+2018}. We simply repeat the main equations here, namely
\begin{align}\label{eq:perturbs1}
\delta'+(1+w)(\theta-3\phi')+3\mathcal{H}(c_s^2-w) \delta &= 0~,\\
\label{eq:perturbs2}
\theta'+\mathcal{H}(1-3c_a^2)\theta-k^2\left(\psi + \frac{c_s^2}{1+w} \delta - \sigma\right) &= 0~,
\end{align}
where local thermal equilibrium sets the sound speed  $c_s^2 = \delta P / \delta \rho$ equal to the \textit{adiabatic} sound speed $c_a^2 = P'/\rho' = w - \frac{w'}{3\mathcal{H}(1+w)}$, and  $w=P/\rho$ is the equation-of-state parameter. We denote derivatives with respect to conformal time by a $'$ symbol, and define $\mathcal{H}=a'/a$, where $a$ is the scale factor. Assuming the $2\to2$ interactions of CanDM are fast, we choose to close the hierarchy by setting $\sigma=0$, as fluids with very fast self-interactions have negligible shear.
\newpage
We show the evolution of $c_s^2$ and $w$ in the left panel of figure \ref{fig:perturbs_eos}, where we observe that CanDM has $c_s^2=w=\tfrac{1}{3}$ in the radiation-like phase, while $w$ and $c_s^2$ fall as $1/\ln a$ in the cannibalistic phase, and tend towards zero as $a^{-2}$ in the cold phase.

In the right panel of figure \ref{fig:perturbs_eos} we additionally show the CanDM perturbations, as well as their effect on the other species through their influence on the total stress-energy tensor and on the metric potentials.
The CanDM perturbations only start growing well into the cold phase. The growth of the CDM and baryon perturbations is suppressed by the presence of the smoother CanDM component, which contributes to the expansion but not to gravitational clustering. This growth suppression mechanism is very similar to that induced by hot or warm dark matter, even if the reason for which CanDM remains smoother than CDM is different.
We can thus expect a suppression in the final matter power spectrum, which we will confirm in section \ref{subsec:pk_implications}.

In the limit of $w,c_s^2 \to 0$, the perturbation \fullmanyeqref{eq:perturbs1} and \eqref{eq:perturbs2} become identical to those of standard CDM. On the other hand, when $c_s^2=w=1/3$, one recovers the perturbation equations of dark radiation, albeit with negligible shear and no higher moments. This implies that during the \textit{relativistic} phase the CanDM species is effectively completely degenerate with an additional dark radiation species characterized through a change in the effective number of relativistic species, \DNeff. If the CanDM particle is massive enough, its transition from \tquote{cannibal} to \tquote{cold} phase takes place during radiation domination, when its energy density is subdominant compared to the radiation, and thus the cannibalistic phase is irrelevant. In this case, the CanDM can effectively be described as warm Dark Matter (WDM), since it still has some decreasing but non-zero thermal temperature $T \propto a^{-2}$ during the stages relevant for the CMB and structure formation.
\section{Cosmological implications}
\label{sec:implications}
\subsection{CMB power spectra}
\label{subsec:cmb_implications}
\begin{figure}[t]
	\centering
	\includegraphics[width=0.7\textwidth]{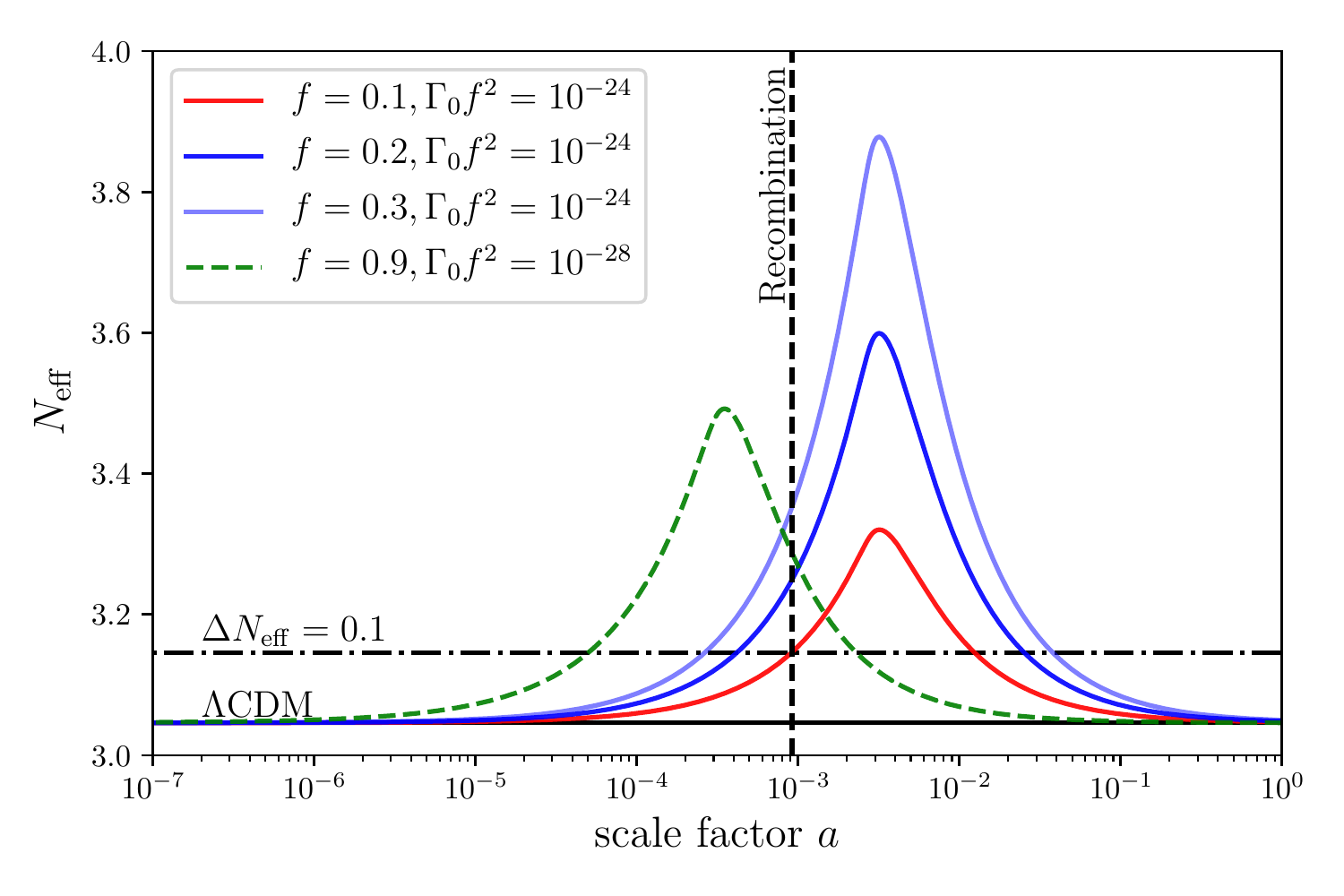}
	\caption{The effective number of relativistic species, $N_\mathrm{eff}$, as a function of scale factor $a$. We show a CanDM model with $f=10\%, m=10\,\mathrm{MeV}, \alpha=10^{9}$, and $n=0$ in red. Increasing $f$ while keeping $\Gamma_0 f^2$ constant increases the height of the peak (blue curves). Changing $\Gamma_0 f^2$ while keeping $f$ constant shifts the position of the peak (green dashed curve).
			We increased the amplitude of the green dashed curve by increasing the fraction to make it more visible. Furthermore, we also plot in black solid the $\Lambda$CDM model ($N_\mathrm{eff}=3.046$), and in black dashed the time of recombination of this model. Finally, we also show a $\Delta N_\mathrm{eff}=0.1$ model as a dot-dashed black line.
		\label{fig:Neff}}
\end{figure}

\begin{figure}[t]
	\centering
	\includegraphics[scale=0.75]{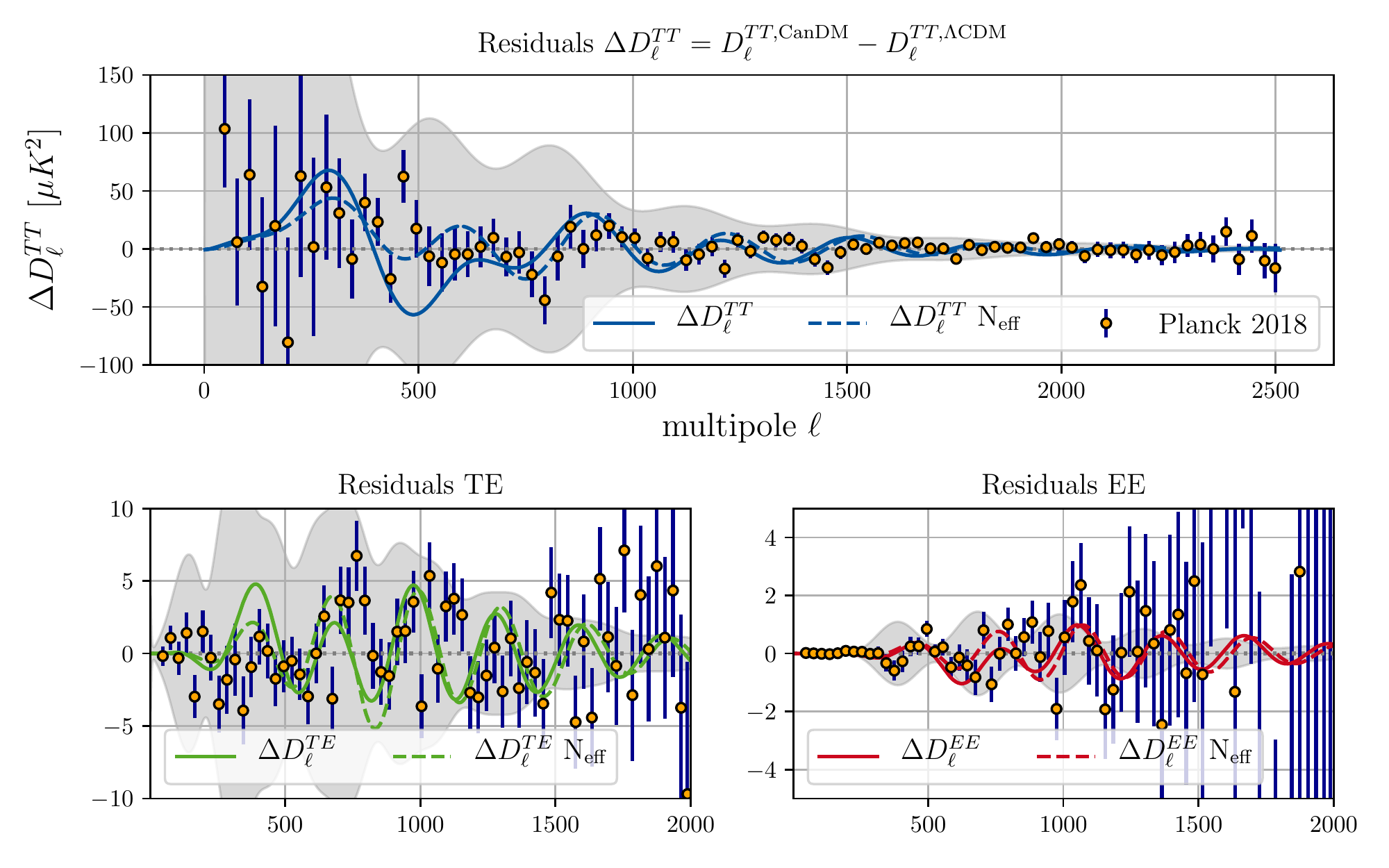}
	\caption{Effects on the CMB power spectra  of CanDM (solid lines) with $f=10\%, m=10\,\mathrm{MeV}, \alpha=10^{9}$, and $n=0$, compared to the impact of  $\Delta N_\mathrm{eff}\approx 0.1$ (dashed lines). 
	We show the residuals $\Delta D_\ell = D_{\ell, \mathrm{CanDM}} - D_{\ell, \mathrm{CDM}}$ or $D_{\ell, \Delta N_\mathrm{eff}} - D_{\ell, \mathrm{CDM}}$ relative to the Planck 2018~\cite{Planck2018} $\Lambda$CDM best-fit model (with one massive and two massless neutrinos), in units of $\mu\mathrm{K}^2$, for each of the TT, TE, EE spectra.
	The corresponding Planck 2018 data point residuals are shown together with their error bars.
	Note that we chose a particularly extreme CanDM model (with $S_8 \approx 0.58$) to emphasize the effect.	}
	\label{fig:delta_Cl_Neff}
\end{figure}

The effects of CanDM on the CMB power spectra are very similar to those of an additional effective relativistic species $\Delta N_\mathrm{eff}$\,. To see why this is the case, we can follow a simple argument: $\Delta N_\mathrm{eff}$ is defined by the equation 
\begin{equation}
	\rho_\mathrm{ultra-relativistic} = \rho_\mathrm{\gamma} \left(1 + \frac{7}{8} \left(\frac{4}{11}\right)^{4/3}  (3.046 + \Delta N_\mathrm{eff})\right)~.
\end{equation}
We thus need to find all ultra-relativistic energy density contributions. Since CanDM is not a pressureless species (unlike CDM), we can approximately split its energy density into a non-relativistic part $\rho-3P$ and an ultra-relativistic part $3P$. 
The ratio of the pressures of photons and CanDM will, therefore, determine the resulting effective number of relativistic species. 

If CanDM is still in its \textit{relativistic} phase at the time of photon decoupling, the effect will be like that of a positive and constant \DNeff due to the CanDM pressure scaling like the photon pressure. If CanDM is already well into its \textit{cold} phase,
the effects on the CMB will be like those of ordinary CDM -- which means, essentially, no effect at all at the perturbation level~\cite{Voruz:2013vqa} -- due to the negligible pressure. Finally, when the CanDM is in the \textit{cannibalistic} phase the phenomenology is richer. Since the pressure of the CanDM only falls as $\sim 1/(a^3 \log^2 a)$ during this phase, and thus slower than the $1/a^4$ scaling of the photons (see the right panel of figure \ref{fig:T_rho_background}), the relative contribution grows. This results in an enhancement of \Neff, which abruptly ends at the beginning of the \textit{cold} phase, where the pressure decreases as $1/a^{5}$ and quickly drops below the photon contribution. In summary, $\Delta N_\mathrm{eff}(z)$ has a peak near the end of the cannibalistic phase. This is important to keep in mind, since we will see in the result section that in the phenomenologically most interesting models, the cannibalistic phase ends roughly near the epoch of recombination.

With the above discussion, we can see that the impact on the CMB will mainly depend on the evolution of $\Delta N_\mathrm{eff}$ before the time of photon decoupling, which in turn depends on the CanDM phase. To illustrate this, in figure \ref{fig:Neff} we compare a CanDM model with $f=10\%$, $m=10\,$MeV, $\alpha=10^{9}$, and $n=0$
(solid line) and a $\Lambda$CDM model with $\Delta N_\mathrm{eff} = +0.1$ (dashed line), which roughly corresponds to the $\Delta N_\mathrm{eff}$ of the considered CanDM model at the redshift of recombination.
As we can see in figure \ref{fig:delta_Cl_Neff}, the resulting imprint of these models on the CMB is very similar. It is not identical though, due to several differences between relativistic CanDM and extra free-streaming particles. Most importantly, at the background level, we have seen than CanDM has a time-varying $\Delta N_\mathrm{eff}(z)$. Besides, at the perturbation level, it behaves like a self-interacting fluid rather than a collection of free-streaming particles, which removes the shear effect and the neutrino drag effect (see e.g.,~\cite{Audren:2014lsa}).

\subsection{Matter power spectrum}
\label{subsec:pk_implications}

\begin{figure}
	\centering
	\includegraphics[scale=0.75]{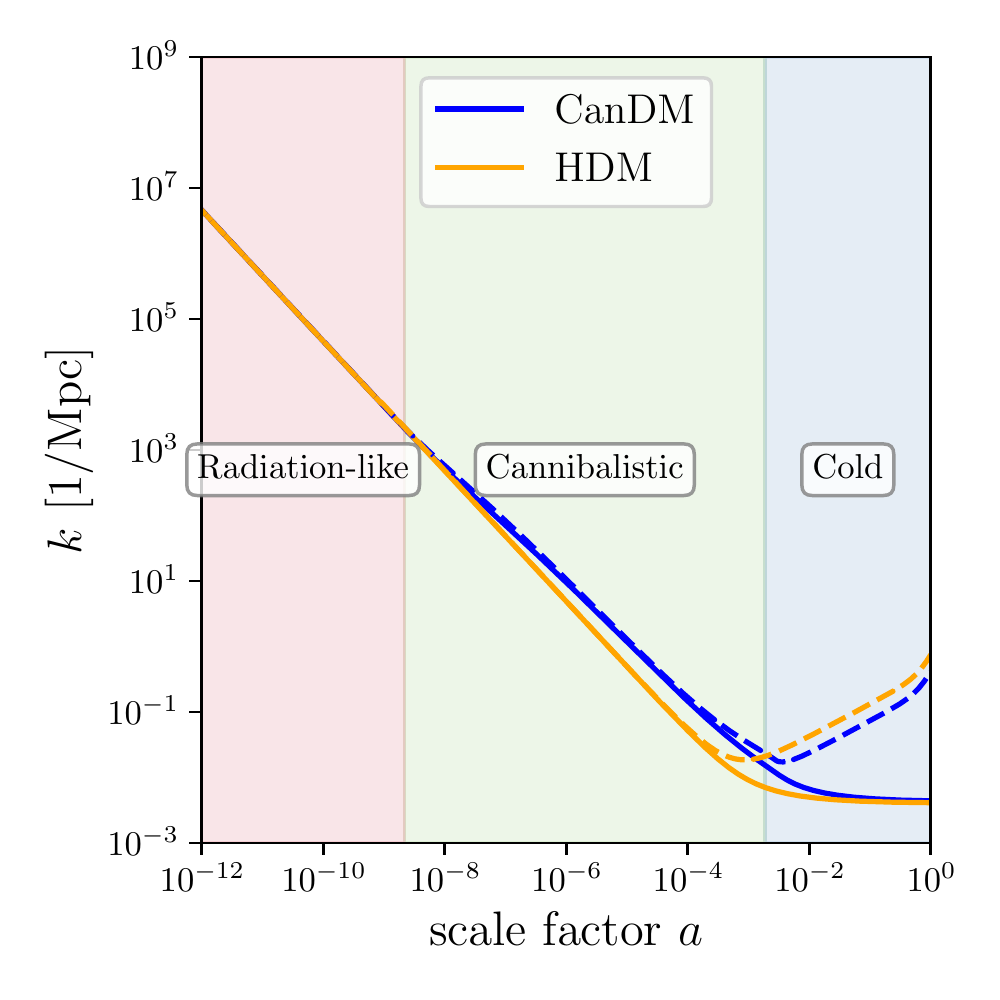}
	\includegraphics[scale=0.75]{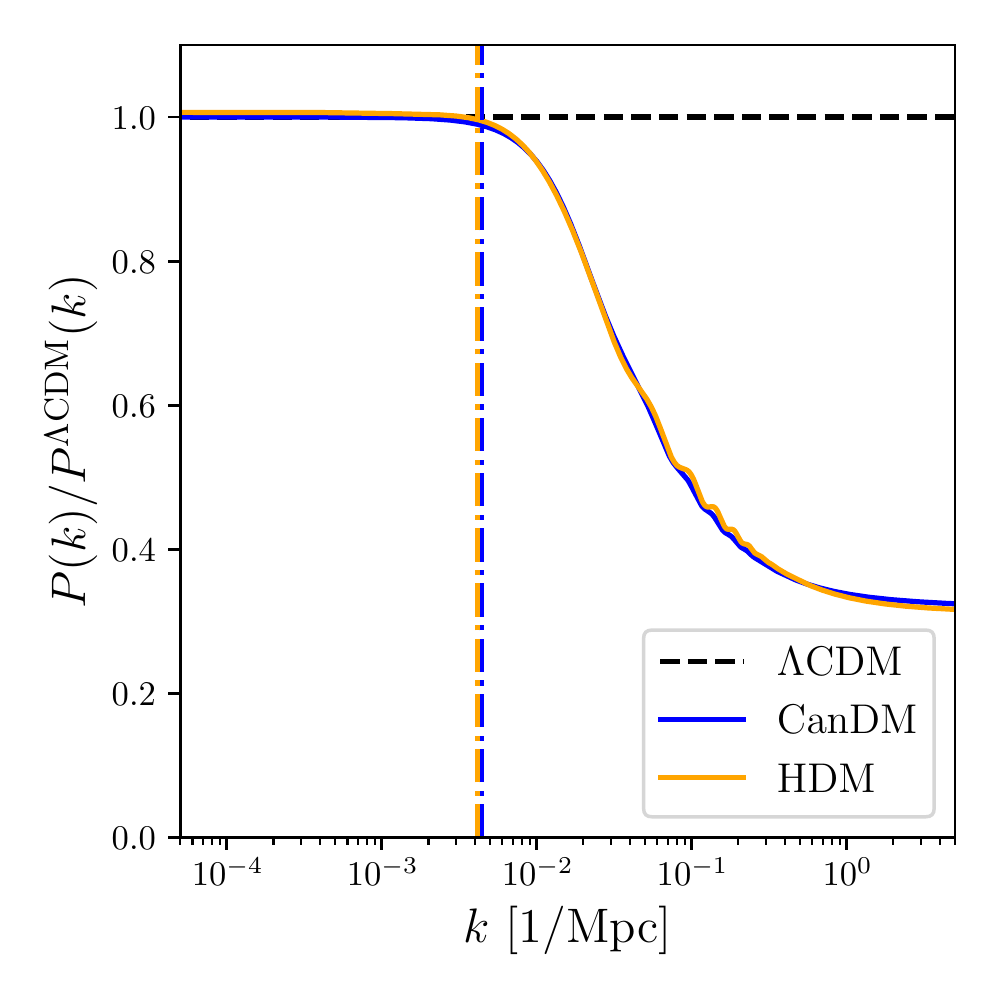}
	\caption{\label{fig:k_scales} \textbf{Left:} The Jeans and free-streaming wavenumbers of two models with 10\% of either CanDM or HDM (dashed lines), and the corresponding horizon wavenumbers (solid lines). The thermal HDM mass was tuned to get the same horizons today (for the same CanDM model as in figure \ref{fig:delta_Cl_Neff}, we adjusted $T_\mathrm{HDM}/m_\mathrm{HDM}$ to $0.70T_\gamma^0/\mathrm{eV}$).
	\textbf{Right:} The power spectrum suppression for the same two models relative to a $\Lambda$CDM model, and the corresponding horizons as vertical dashed lines.
		\label{fig:jeans_and_pk}
}
\end{figure}

As CanDM has a non-negligible pressure, we expect it to suppress the matter power spectrum on small scales, similarly to dark matter particles with a significant velocity dispersion. We actually find that in the range of parameters in which we are interested, each CanDM model has an equivalent WDM model in terms of the power spectrum suppression, with a given value of the WDM velocity dispersion -- or equivalently of the ratio $T_\mathrm{WDM}^0/m_\mathrm{WDM}$. The most interesting CanDM models from the point of view of reducing the $S_8$ tension can be mapped to models with a very large velocity dispersion (just a few orders of magnitude below that of active neutrinos) that should rather be called HDM models. We provide an example of such an equivalence at the level of the matter power spectrum in figure~\ref{fig:k_scales} (right panel).

\interfootnotelinepenalty=10000
In more details, the effects of CanDM on structure formation
depend mainly on the evolution of the density perturbations. By combining \fullmanyeqref{eq:perturbs1} and \eqref{eq:perturbs2}, we obtain a differential equation with frequency\footnote{This standard result is displayed in analogy to literature on WDM. In the limit in which CanDM dominates the universe density, one gets an extra contribution $\frac{3}{2}{\cal H}w/a^2$, which explains the small difference between this approximate result and the equivalent one in \cite{Erickcek:2020wzd}. We have explicitly checked that figure \ref{fig:jeans_and_pk} looks the same for both definitions.}
${\omega^2 = \left(\frac{3}{2} \mathcal{H}^2 - k^2 c_s^2\right)/a^2}$. 
This leads us to define the Jeans wavenumber $k_J(a) = \sqrt{3/2}\, \mathcal{H}/c_s$ such that $\omega^2>0$ if $k < k_J(a)$ and the solution remains oscillatory. On the other hand if $k > k_J(a)$ the solution is allowed to grow.
This so-called Jeans instability is a result of the CanDM thermal self-interactions preventing collapse below the Jeans scale $\lambda_J = 2\pi/k_J(a)$. 
For WDM or HDM there is a similar suppression mechanism caused by the thermal velocity dispersion of the species. In this case, the suppression takes place for wavenumbers bigger than the free-streaming wavenumber
$k_\mathrm{FS}=\sqrt{3/2}\, \mathcal{H}/c_s$ where $c_s$ is the WDM or HDM sound speed. 

We show these functions in the left panel of figure \ref{fig:k_scales}, together with the corresponding horizon wavenumbers defined as $1/k_H(a)=\int_0^a 1/k_J(\tilde{a})\,\,\textrm{d}\ln \tilde{a}$. 
While the Jeans and free-streaming lengths tell us which scales are damped by pressure \textit{at a given time}, the corresponding horizons represent the distance travelled by a sound wave between the early universe and this time, and indicate up to which scale the growth of structure can be affected by pressure effects \cite{Boyarsky:2008xj}.
In the right panel of figure \ref{fig:k_scales}  we show the resulting suppression of the matter power spectrum for these models relative to an equivalent \lcdm model.
We can check that the suppression starts on a wavenumber corresponding to the Jeans (or free-streaming) {\it horizon} today. It saturates above a wavenumber corresponding to the Jeans (or free-streaming) {\it length} today, since such scales have not begun to collapse gravitationally.

\begin{figure}[t]
	\centering
	\includegraphics[scale=0.75]{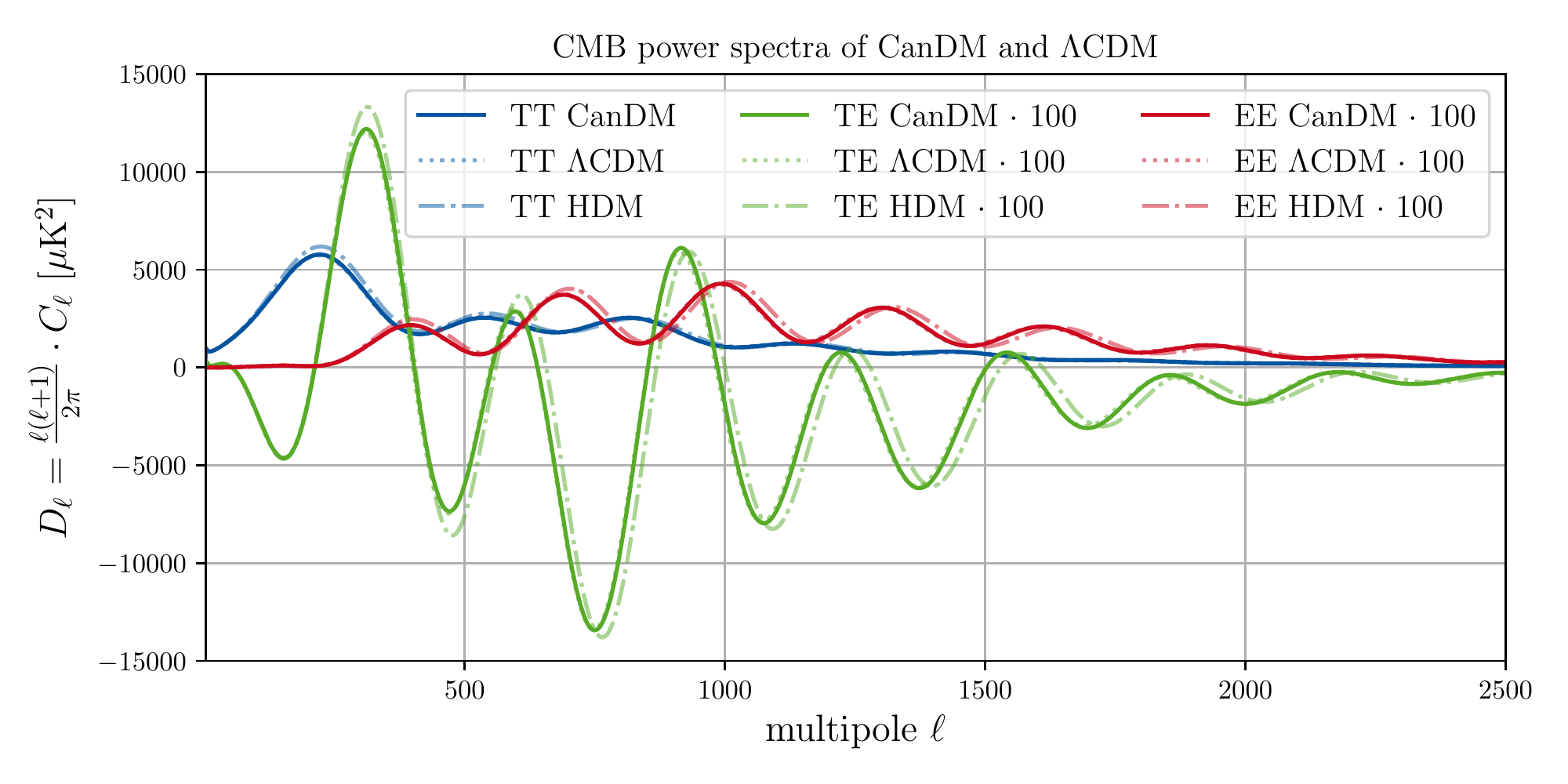}
	\caption{CMB TT, TE, EE spectra for the same models as in figure~\ref{fig:jeans_and_pk}, with 10\% of either CanDM (solid lines) or HDM (dot-dashed lines), and matching values of the Jeans and free-streaming horizon wavenumbers. The corresponding plain $\Lambda$CDM model is plotted in dotted lines, but remains almost indistinguishable from the CanDM case: this shows that CanDM can have the same strong impact on the matter power spectrum as HDM, while, unlike the latter,  leaving the CMB spectrum almost unaffected.}
	\label{fig:Cl_WDM}
\end{figure}

It is worth stressing that the mapping between CanDM models and equivalent light WDM or HDM models holds only at the level of the matter power spectrum, and not of the CMB spectrum. The reason is that  in the CanDM model, the non-relativistic transition takes place at the beginning of the cannibalistic phase. During this phase, the sound speed slowly decreases (see the left panel in figure~\ref{fig:perturbs_eos}). Thus it gets reduced compared to that of WDM or HDM, which would remain close to $1/\sqrt{3}$ up to the non-relativistic transition. This means that CanDM behaves momentarily in a \enquote{colder way} than its free-streaming counterpart. Indeed, its Jeans wavenumber is temporarily larger, as can be seen in the left panel of figure~\ref{fig:jeans_and_pk}, and thus more scales cluster like CDM. In the equivalent WDM or HDM model, the same scales would experience damped oscillations, which would couple gravitationally to photon perturbations and change the amplitude and phase of CMB acoustic oscillations. Additionally, just after the time of photon decoupling, they would induce an Integrated Sachs-Wolfe effect. Thus, as long as photon decoupling takes place during or after the cannibalistic phase, the CanDM model has less impact on CMB spectra than its light WDM or HDM counterpart, as can be checked in figure~\ref{fig:Cl_WDM}. This makes CanDM a particularly interesting candidate to address the $S_8$ tension while remaining compatible with high precision CMB data.

The amplitude of the step-like suppression of the matter power spectrum induced by CanDM in mixed CDM+CanDM models depends only on $f$ and could be estimated in the same way as for CDM+WDM models. In the CDM+HDM limit, the suppression factor is given given by $(1-8f+{\cal O}(f^2))$ like for ordinary massive neutrino models. However, if we want to understand which CanDM model could help solving the $S_8$ tension, it is more interesting to estimate the {\it wavenumber} at which the suppression starts as a function of CanDM parameters. For this, we should in principle compute the Jeans horizon, but it is easier to estimate the maximum Jeans scale (or minimum Jeans wavenumber), which is of the same order of magnitude as the Jeans horizon for light WDM or HDM models (see the left plot in figure~\ref{fig:jeans_and_pk}).
\enlargethispage{1\baselineskip}

Given that the Jeans scale $\lambda_J$ has a maximum around the beginning of the \textit{cold} phase of CanDM, we would like to find a simple analytical
approximation for this time, which will help us understand our results more deeply. For this we can equate the interaction rate $\Gamma_{3\to2}$ to the Hubble rate $H$.
In this work we consider only cases where the
transition happens in the non-relativistic (NR) phase. Thus we can get $\Gamma_{3\to 2}$ from equation \eqref{eq:gamma32} with $n_\mathrm{NR}=\rho/m$ and infer the scale factor of the transition, $a_\mathrm{cold}$, from
\begin{gather}
H(a_\mathrm{cold}) = \frac{\alpha^3}{m^7} \rho^2_\can \left(\frac{T}{T_\mathrm{*}}\right)^n
\\\iff
\underbrace{\frac{H(a_\mathrm{cold})}{H_0 \sqrt{\Omega_r}}\cdot a_\mathrm{cold}^6}_{\approx a_\mathrm{cold}^4\mathrm{\ for\ RD}} = 
\underbrace{\vphantom{\left(\frac{\alpha^3}{m^{7-n}}\right)}\frac{\alpha^3}{m^{7-n}}}_{\Gamma_n}
\frac{f^2}{T_*^n} 
\underbrace{\frac{(\Omega_m-\Omega_b)^2\rho^2_\mathrm{crit, 0}}{H_0 \sqrt{\Omega_r}}}_{\approx (69\,\mathrm{eV})^7} 
\underbrace{\left(\frac{T}{m}\right)^n}_{\mathcal{O}(0.1^n)}~.
\label{eq:Gammafactor}
\end{gather}
If the transition happens before matter-radiation equality ($a_\mathrm{cold}\ll a_\mathrm{eq}$, 
true for most models), the left-hand side of equation \eqref{eq:Gammafactor} simplifies to just $\mathrm{a^4_\mathrm{cold}}$. On the right-hand side, the second ratio depends only on the $\Lambda$CDM parameters, and keeps roughly the same value for viable models close to the Planck best fit.  The ratio $T/m$ is always of order one when the cannibalistic phase begins and only drops logarithmically before the cold phase: thus $(T/m)^{1/4}$ has a very moderate impact on $a_\mathrm{cold}$. We can conclude that the value of $a_\mathrm{cold}$ is mainly controlled by the combination $\Gamma_n f^2 T_*^{-n}$, where we have defined $\Gamma_n \equiv \alpha^3 m^{n-7}$. Since $T_*$ has only a weak dependence on the other CanDM parameters, we omit it in the definition of $\Gamma_n$.
This discussion gives us a physical intuition on why our results should mainly
	depend on $\Gamma_n f^2$, which we will confirm in the following sections.

\begin{figure}[t]
	\centering
	\includegraphics[scale=0.75]{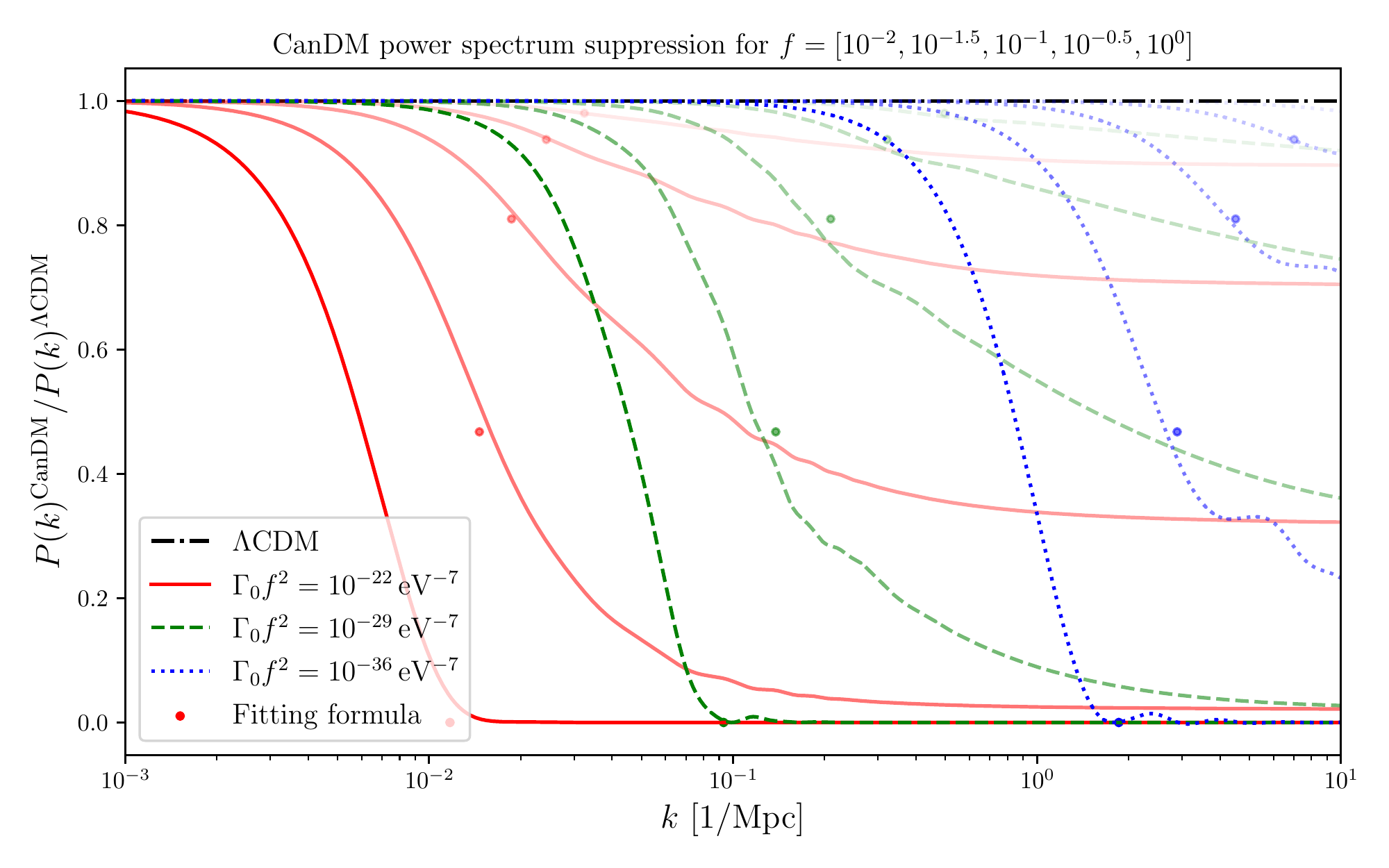}
	\caption{The matter power spectrum for different CanDM models relative to the $\Lambda$CDM model. The colors and line-styles show three
		different masses $m=10^4$\,eV (solid red), $10^5$\,eV (dashed green), and $10^6$\,eV (dotted blue) with $\alpha=10^2$. The opacity of the line indicates the fraction of CanDM from 1\% to 100\% and the dots represent the wavenumber of the inflection point, $k_\mathrm{step}$\,, expected from equation \eqref{eq:fitting_formula_suppression}.
		\label{fig:Pk_suppression}}
\end{figure}

Having found the time of the \textit{cold} transition, one can also find the corresponding\footnote{This is the wavenumber that crosses the Hubble horizon exactly at the given scale factor.} wavenumber $k_\mathrm{cold} = a_\mathrm{cold}H(a_\mathrm{cold})$, which mostly determines the characteristic scale of the step 
in the matter power spectrum. We found empirically that the inflection point of this step is given by
\begin{equation}
	k_\mathrm{step} \approx (1.48\pm 0.03) \left(\frac{k_\mathrm{cold}}{\mathrm{Mpc}^{-1}}\right)^{(0.77 \pm 0.01)} \mathrm{Mpc}^{-1} \,.
	\label{eq:fitting_formula_suppression}
\end{equation}
Figure \ref{fig:Pk_suppression} confirms that the power spectrum suppression is mostly determined by $\Gamma_n f^2$ (scale of the step) and $f$ (amplitude of the step). The dots show the values of $k_\mathrm{step}$ given by equation~\eqref{eq:fitting_formula_suppression}. 
We see that an increase of $\Gamma_n f^2$ leads directly to a decrease of $k_\mathrm{step}$ and a suppression on larger scales.
We note that similar considerations for the case of a decaying cannibalistic species can be found in \cite{Erickcek:2020wzd}. In that work, the discussion is also extended to the formation on microhalos, which are not considered here. 

We can estimate the expected order of magnitude of the parameter constraints for $\Gamma_n$ as follows: CMB data requires CanDM to be sufficiently non-relativistic at the time of photon decoupling, such that it behaves like a cold dark matter species around this time, leaving barely any impact on the CMB. As we will later see, current CMB data is sensitive enough to impose approximately $T/m < 10^{-5}$ at decoupling. At the end of the \textit{relativistic} phase, the temperature will generally be of order $T/m \sim 1$ since this is necessary for the inefficiency of the $2 \to 3$ interaction. Afterwards, $T/m$ only falls logarithmically, such that we can expect it to be around $T/m \sim 0.1$ at the end of the \textit{cannibalistic} phase, which is precisely the time $a_\mathrm{cold}$ computed above. Finally, the temperature will drop as $a^{-2}$ in the ensuing \textit{non-relativistic} phase. To let $T/m$ decrease until $10^{-5}$ starting from about 0.1 at $a_\mathrm{cold}$, we need to wait until the scale factor has increased by two orders of magnitude. Thus we expect that CanDM does not spoil the CMB spectra under the condition $100 a_\mathrm{cold} < a_\mathrm{dec}$, that is, $ a_\mathrm{cold} < 1/100 \cdot 1/1080 \approx 10^{-5}$. If we insert this condition into equation \eqref{eq:Gammafactor} and assume $n=0$, we get $\Gamma_0 \sim 10^{-33}\, \mathrm{eV}^{-7}$. For a time-dependent velocity-averaged cross section, since $a_\mathrm{cold}$ depends on $n$ as $0.1^n \Gamma_n T_*^n$ we can translate this into an estimate for $\Gamma_2$ using $T_* \sim 1\,\mathrm{GeV}$ (justified from figure~\ref{fig:T_rho_background}) and $n=2$. We thus estimate an upper bound $\Gamma_2 \sim 10^{-13}\,\mathrm{eV}^{-5}$. We will see in section \ref{sec:results} that these estimated bounds agree well with the actual bounds.

In conclusion to this section, we see that CanDM combines only the advantageous effects of the 
$\Delta N_\mathrm{eff}$ and WDM (or HDM) models. It induces a suppression in the matter power spectrum, similarly to light WDM or HDM, that may ease the $S_8$ tension. It avoids, however, the strong impact of the latter models on the CMB spectra that would contradict Planck data (see figures \ref{fig:jeans_and_pk} and \ref{fig:Cl_WDM}). CanDM can still manifest itself in the CMB spectrum in a similar way as a positive $\Delta N_\mathrm{eff}$ (see figure \ref{fig:delta_Cl_Neff}), which might ease the Hubble tension, but with a different consequence for the matter power spectrum. To check whether this model can really solve tensions while remaining compatible with observations, we need to confront it to the actual data likelihoods.
\section{Results}\label{sec:results}
In this section we use our implementation of CanDM in \class to constrain this model with various cosmological probes detailed below. 
In section \ref{subsec:chains} we show the parameter constrains from CMB and BAO measurements
and in section \ref{subsec:sigma8chi2}
we investigate the effect of this model on the $S_8$ tension with weak lensing measurements.

\subsection{Parameter constraints}
\label{subsec:chains}
We use the Planck 2018 temperature, polarization, and lensing likelihoods\footnote{This includes the {low-$\ell$ TT}, {low-$\ell$ EE}, {high-$\ell$ TTTEEE}, and {lensing} likelihoods.} as described in \cite{Aghanim:2019ame} and \cite{Aghanim:2018oex}. We add the BAO data from data release 12 of SDSS, \cite{Alam:2016hwk} as well as that of the 6DF survey \cite{Beutler_2011}, and the main galaxy sample of data release 7 of SDSS \cite{Ross:2014qpa}. We henceforth refer to this combination as P18+lens+BAO.

We make use of the MCMC sampler \textsc{MontePython-3} \cite{Audren:2012wb, Brinckmann:2018cvx} to perform a Bayesian parameter exploration on the chosen cosmological parameter basis.
We consider the Planck baseline model with two massless neutrinos and one massive neutrino with mass $0.06\,$eV, replacing a fraction $f$ of the CDM with CanDM. We thus use the standard \lcdm parameters \{$\omega_b$, $\omega_\mathrm{cdm}$, $H_0$, $\ln (10^{10} A_s)$, $n_s$, $\tau_\mathrm{reio}$\} as well as the additional CanDM parameters \{$m$, $f$, $\alpha$ or $\Gamma_n$\} with a fixed temperature dependence of the thermally averaged cross section, either $n=0$ or $n=2$. We always assume top-hat priors on $\lnM$ and either $\lnalph$ or $\log_{10}\Gamma_n$, and the fraction $f$ is either varied, fixed to 100\%, or fixed to 1\% (a representative example of sub-dominant CanDM chosen in \cite{BuenAbad+2018}).

\begin{figure}[t]
 \centering
 \includegraphics[width=0.49\textwidth]{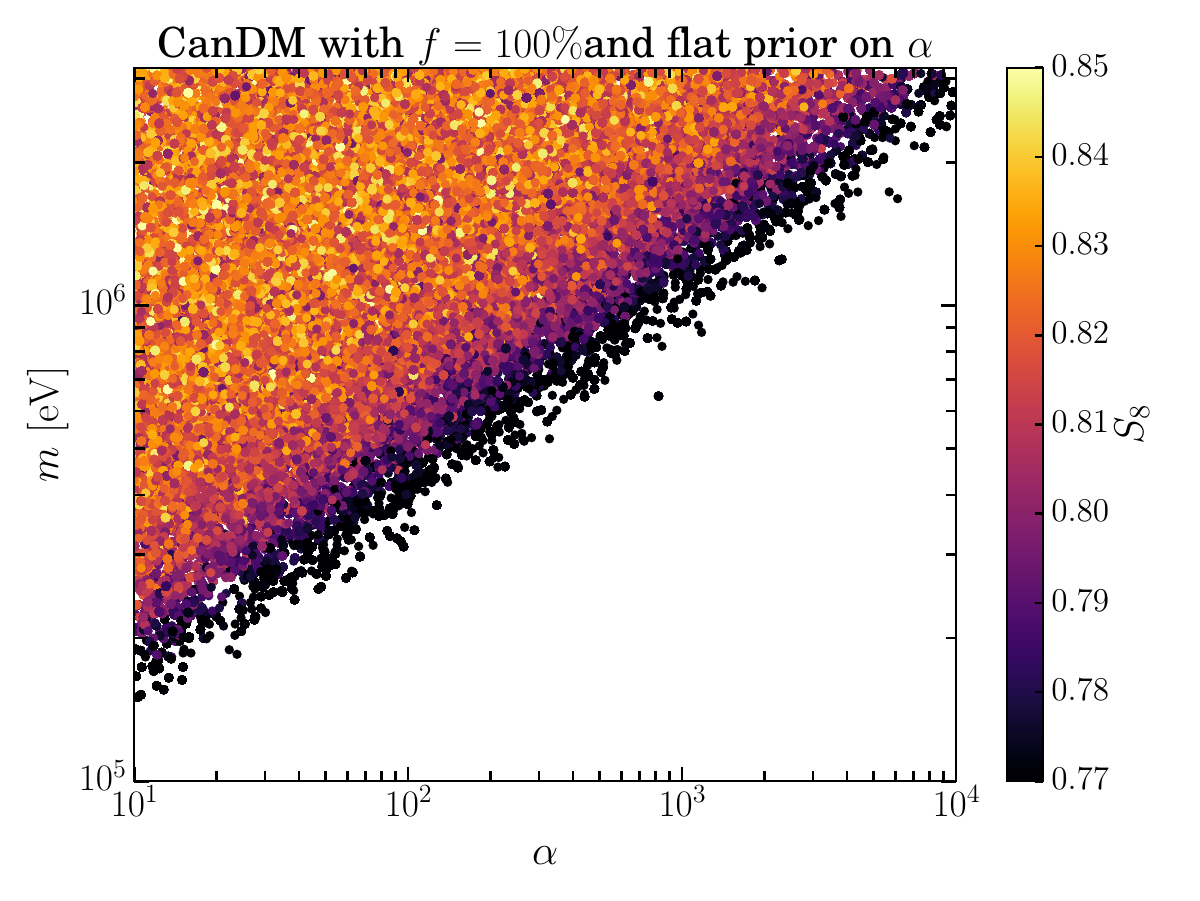}
 \includegraphics[width=0.49\textwidth]{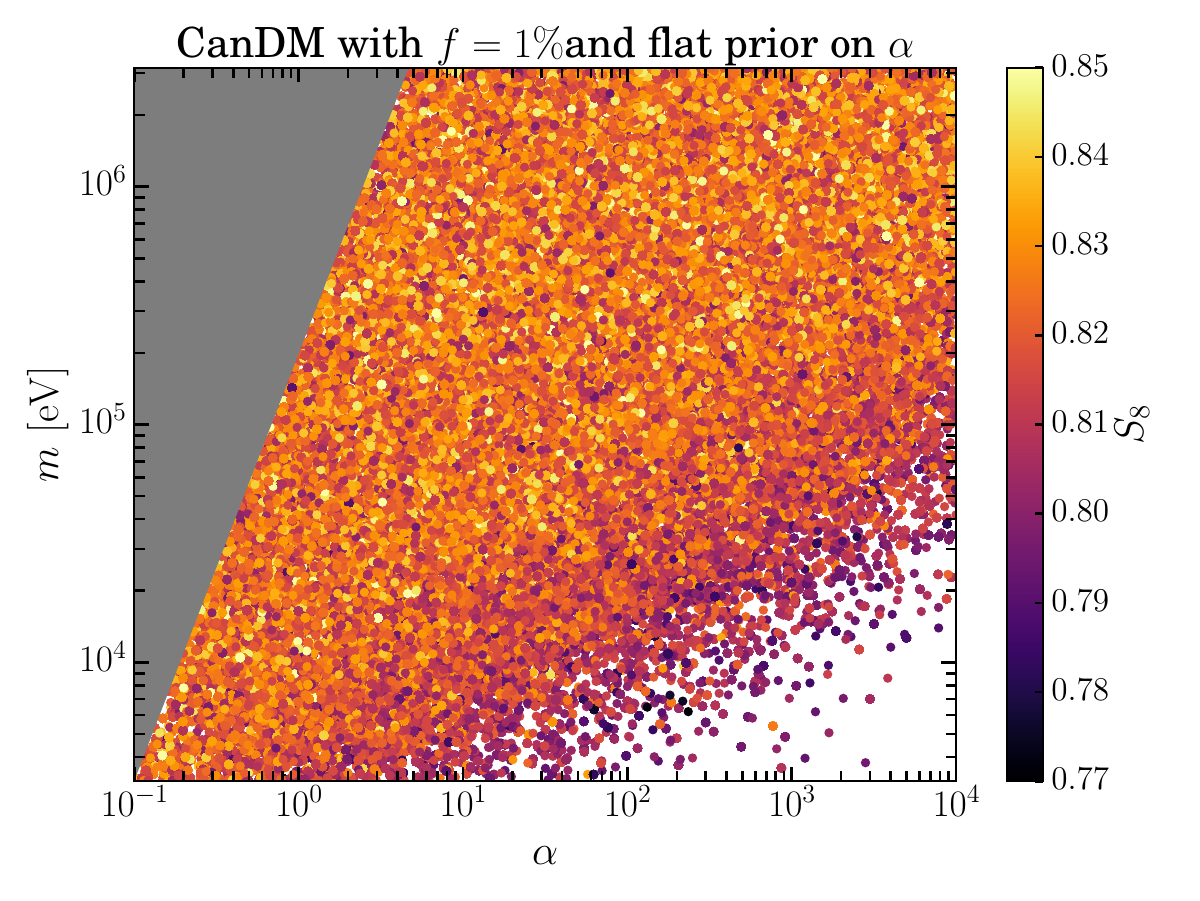}
 \caption{For models with a fixed CanDM fraction $f$ we show the distribution of $S_8$ values as a function of the mass \mcandm~and the interaction parameter \alphcandm, as constrained by CMB + BAO data. In solid gray we show the region where the assumption of non-relativistic evolution cannot be made when the species exits chemical equilibrium (see text for more details).  \textbf{Left:} For the case of $f=100\,\%$. \textbf{Right:} For the case of $f=1\,\%$.}
 \label{fig:M_a_fullfrac}
\end{figure}

\begin{figure}[t]
 \centering
 \includegraphics[width=0.49\textwidth]{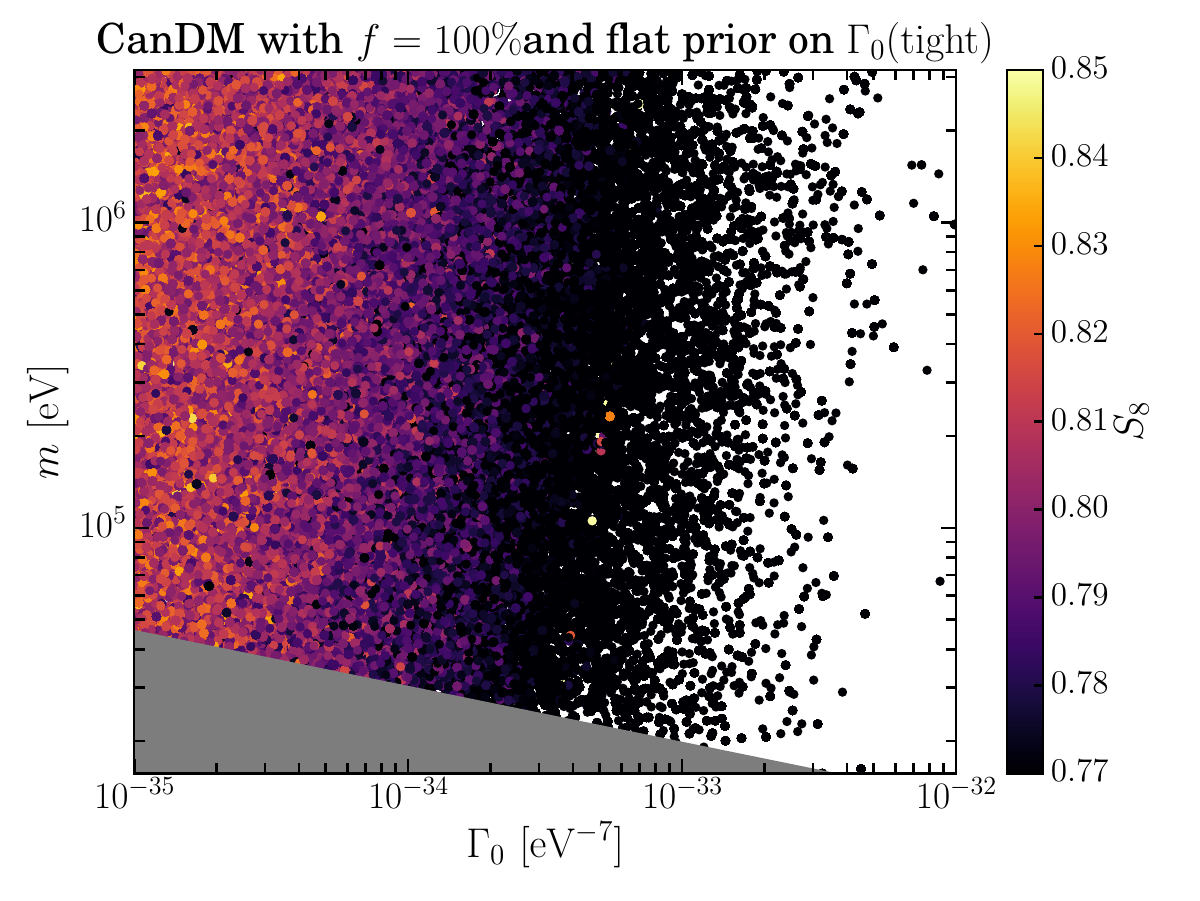}
 \includegraphics[width=0.49\textwidth]{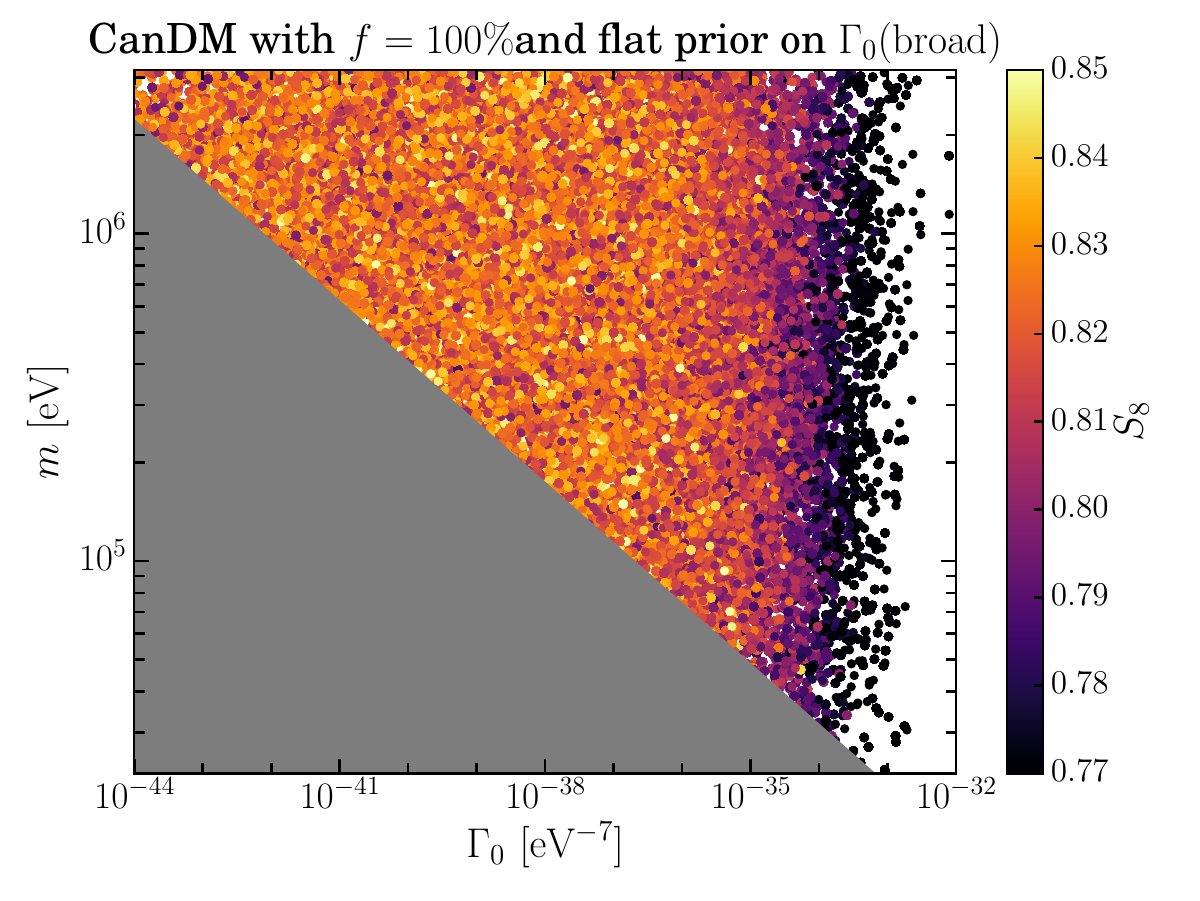}
 \caption{For CanDM representing all of the dark matter, we show the distribution of $S_8$ values as a function of the mass $m$ and the interaction parameter $\Gamma_0 = \alpha^3/m^7$\,. Both plots have a logarithmic prior on $\alpha^3/m^7$, but in different ranges. In solid gray we show the region where the assumption of non-relativistic evolution cannot be made when the species exits chemical equilibrium (see text for more details). \textbf{Left:} Tight prior of [-35,-25]. \textbf{Right:} Broad prior of [-45,-25].}
 \label{fig:m_gamma_fullfrac}
\end{figure}

\begin{figure}[t]
 \centering
 \includegraphics[width=0.7\textwidth]{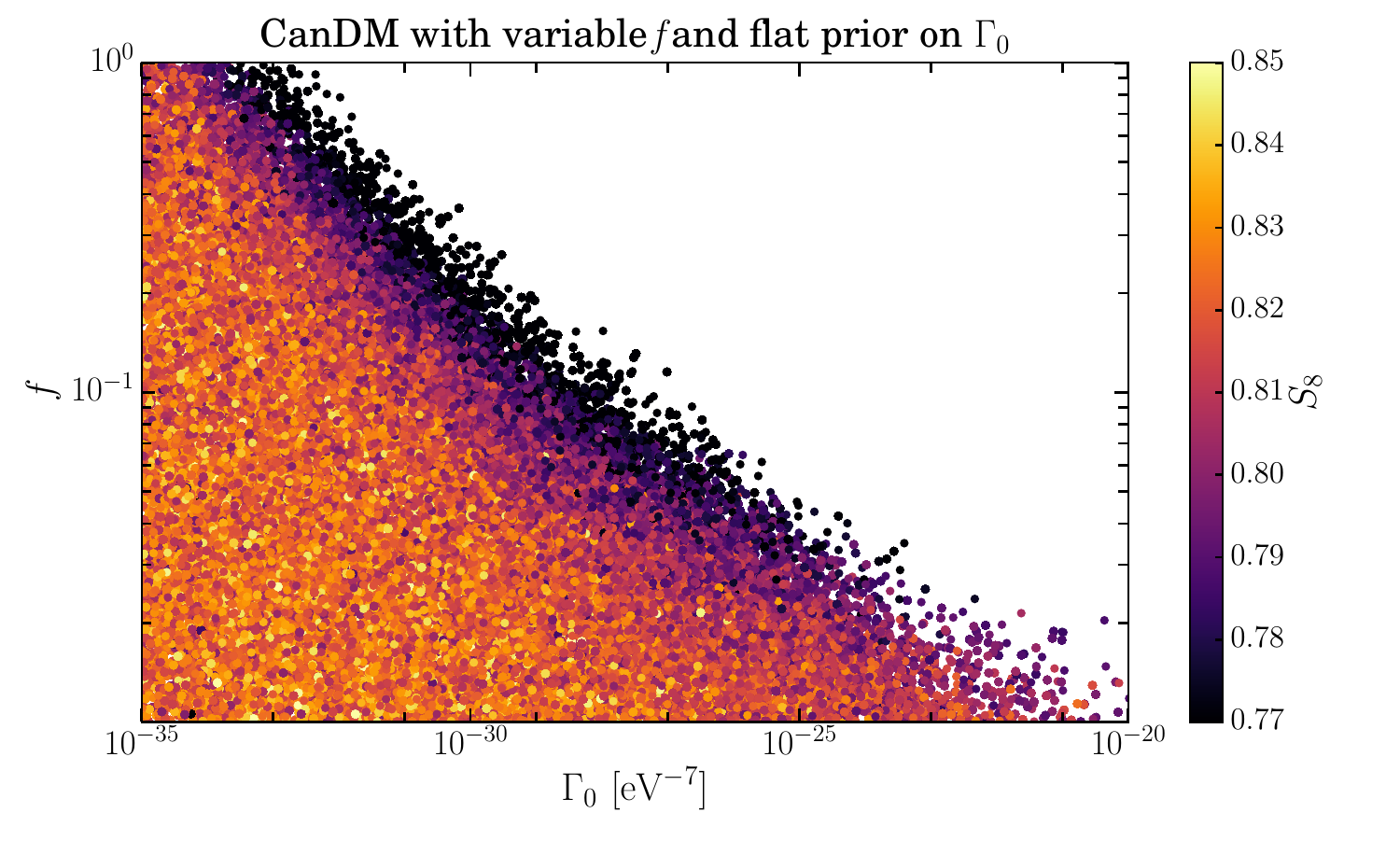}
 \caption{For a varying fraction we show the distribution of $S_8$ values as a function of the fraction \fcandm\ and the interaction parameter $\Gamma_0$\,, marginalized over all possible masses. Where the $S_8$ value is lower, the model is naturally excluded by Planck+BAO data.}
 \label{fig:f_a_varfrac}
\end{figure}

\begin{figure}[t]
 \centering
 \includegraphics[width=0.49\textwidth]{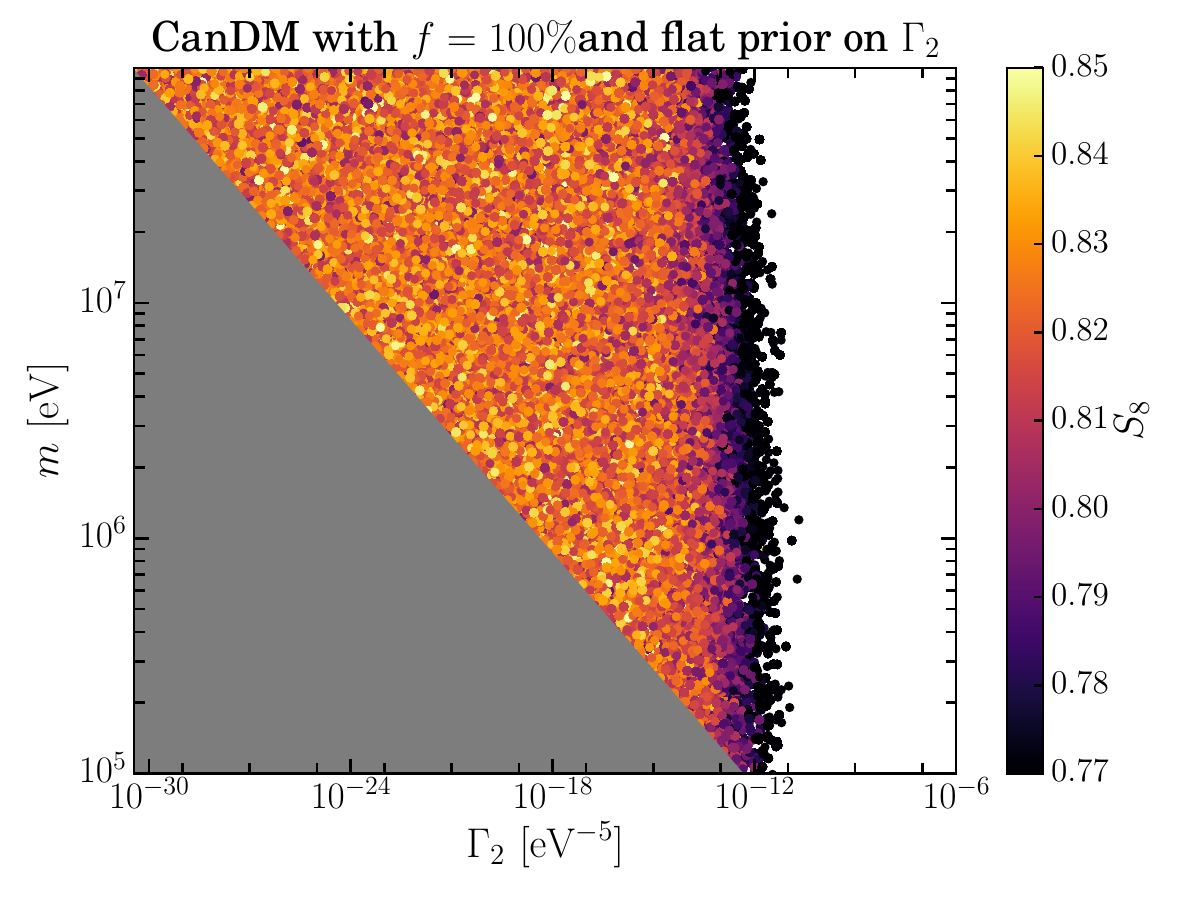}
 \includegraphics[width=0.49\textwidth]{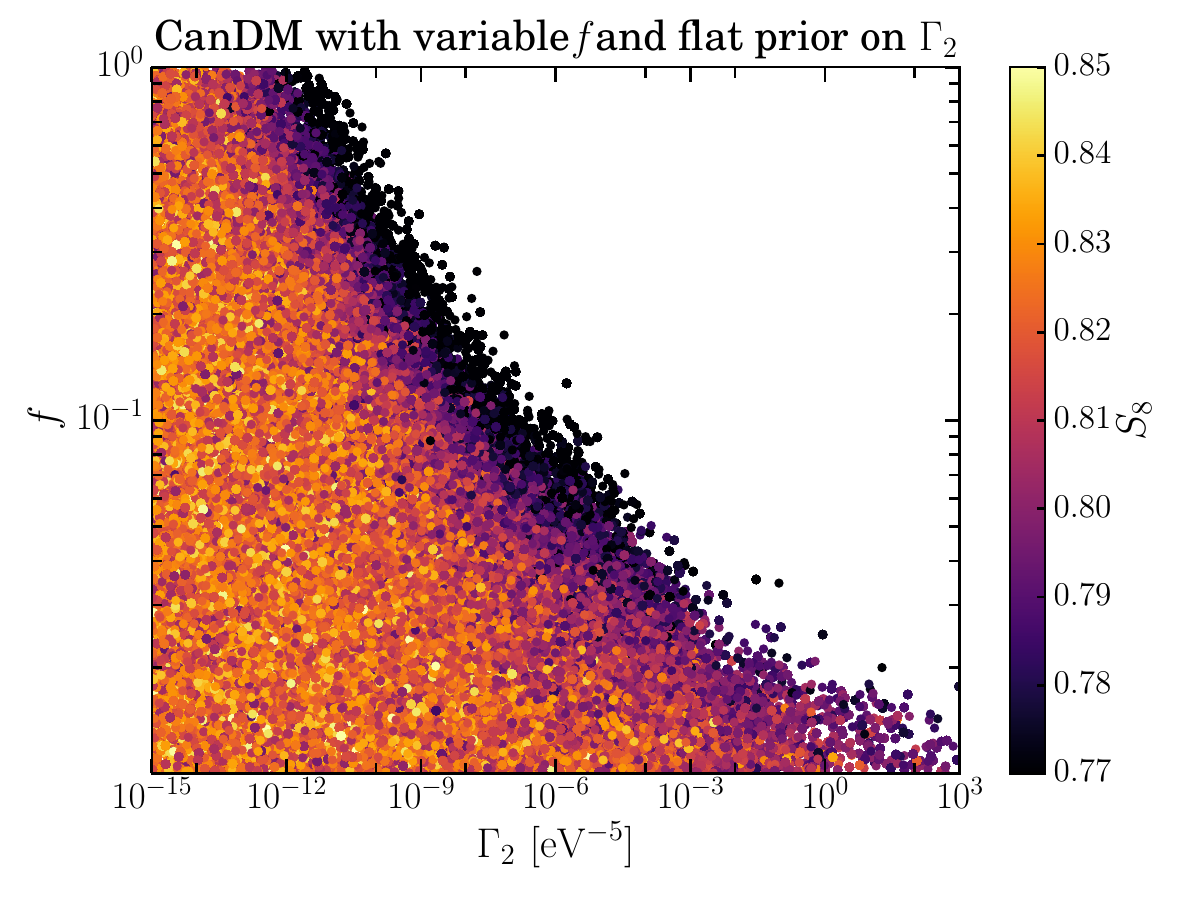}
 \caption{We show the distribution of $S_8$ values as a function of the mass $m$ and the interaction parameter $\Gamma_2 = \alpha^3/m^5$\, for a model with $\Gamma_{3\to 2} \propto T^2$. In solid gray we show the region where the assumption of non-relativistic evolution cannot be made when the species exits chemical equilibrium (see text for more details). \textbf{Left:} CanDM represents all of dark matter. \textbf{Right:} CanDM represents a variable fraction $f$ of dark matter.}
 \label{fig:T2_m_gamma_full_and_varfrac}
\end{figure}

We display the results of our MCMC analysis using the P18+lens+BAO dataset in \cref{fig:M_a_fullfrac,fig:m_gamma_fullfrac,fig:f_a_varfrac,fig:T2_m_gamma_full_and_varfrac}. 
Each dot represents a model in the chains, and the dots are colored according to their $S_8$ value.
The points in the MCMCs are distributed according to the posterior distribution (apart from multiplicities), but for an easy visualization of the 95\% confidence limits, we only show here the
points within 2$\sigma$ of the bestfit\footnote{This means we plot all points with $\Delta\chi^2<\Delta\chi^2_{2\sigma}$. This latter value is given by the solution of $F(\Delta\chi^2_{2\sigma},N_\mathrm{param}) = \mathrm{erf}(2/\sqrt{2}) \approx 95.4\%$ where $F(x,k)$ is the $\chi^2$ cumulative distribution function of $k$ degrees of freedom and $N_\mathrm{param}$ is the total number of parameters for the MCMCs.}. We show only the additional parameters in these figures because the standard $\Lambda$CDM parameters (including $\Omega_m$) do not change with respect to the $\Lambda$CDM model. This is also the reason why we do not include additional supernova data.

In the region of high mass $m$ and small coupling $\alpha$ the assumption that chemical
equilibrium ends during the non-relativistic regime 
is no longer valid. Then, no switching point (green dashed line in figure \ref{fig:T_rho_background})
can be found and, therefore, our method of computation (described in section \ref{subsec:bgevo}) cannot be applied. These regions are marked with gray wedges in all of the relevant figures.

We consider first the case of a temperature-independent thermally averaged cross section $\sigmavsq$ , corresponding to $n=0$. We initially adopt a flat prior on $\lnM$ and $\lnalph$. For the purpose of investigating the phenomenology of the contours, we first discuss the results obtained with a fixed CanDM fraction, $f=100\%$ or $f=1\%$, displayed in figure \ref{fig:M_a_fullfrac}. There are four main observations that we can take away from this figure:
\newpage
\begin{enumerate}
 \item There are points sampled with values of $S_8$ low enough to alleviate the $S_8$ tension (the central value from \cite{Joudaki:2019pmv} is $S_8\simeq0.76$), but these points lie close to the edges of the region preferred by P18+lens+BAO. Indeed, in order to significantly reduce $S_8$, the CanDM model requires the matter power spectrum to be suppressed beyond relatively small values of $k_\mathrm{step}$ of the order of 0.2h/Mpc. This is only possible with large enough values of $\Gamma_n f^2$ (see figure \ref{fig:Pk_suppression}). In turn, this will require a large $a_\mathrm{cold}$ (see equation \eqref{eq:Gammafactor}) and a late transition into the cold regime. However, together with the requirement of generating a certain cosmological abundance today, this implies that the CanDM will have a non-negligible contribution as an early additional light degree of freedom, which is strongly bounded by CMB, BBN, and BAO \cite{Schoneberg:2019wmt} data. This is precisely why even larger values of $\Gamma_n f^2$ and smaller values of $S_8$ are excluded by our P18+lens+BAO dataset, and why all the models solving the $S_8$ tension lie at the edge of the preferred region.
 \item The parameter region we cannot compute (gray wedge at low $\alpha$ and high $m$) does not impact this conclusion,
  since it lies in the region of high $S_8$ values.
 \item The impact on $S_8$ only depends on $\Gamma_0 \equiv \alpha^3/m^7 = \Gamma_{3\to 2}/\rnr^2$. For high masses, this corresponds to lines of constant non-relativistic transition (see section \ref{subsec:pk_implications}), which is the main factor in the late-time evolution of the CanDM species. 
 \item This result holds for each fixed fraction individually. The $f=100\%$ model has the biggest influence on $S_8$ under these constraints, and lower fractions have decreasing capability of changing $S_8$ while remaining compatible with the Planck data.
\end{enumerate}
In particular, the third observation motivates using $\Gamma_0$ to parameterize the interaction
instead of $\alpha$. With this in mind, in figure \ref{fig:m_gamma_fullfrac} we show our results as a function of $\Gamma_0=\alpha^3/m^7$.

Figure \ref{fig:m_gamma_fullfrac} also illustrates the impact of the choice of priors, which are all summarized  in Table \ref{tab:priors}. The narrower prior on $\Gamma_0$ (left plot) gives more emphasis to the region in parameter space where $S_8$ is reduced and well compatible with weak lensing measurements.

\begin{table}
\begin{tabular}{lllcc}
 \toprule
 Figure&Name & Configuration& $\lnM$ & $\log_{10}\Gamma_n$ \\
 \midrule
 \ref{fig:M_a_fullfrac} (left) & f=1 $\alpha$ & $f=100\%$ and a flat prior on $\alpha$ &   [3.5,\,6.5]  &  - \\
 \ref{fig:M_a_fullfrac} (right) & f=0.01 $\alpha$ & $f=~~~1\%$ and a flat prior on $\alpha$ &   [3.5,\,6.5]  &  - \\
 \ref{fig:m_gamma_fullfrac} (left) & f=1  $\Gamma_0$ (tight) & $f=100\%$ and a tight flat prior on $\Gamma_0$ &  [3.5,~~~6] & [-35,\,-25]\\
 \ref{fig:m_gamma_fullfrac} (right)& f=1  $\Gamma_0$ (broad) & $f=100\%$ and a broad flat prior on $\Gamma_0$& [3.5,~~~6] &  [-45,\,-25]\\
 \ref{fig:f_a_varfrac} & variable f $\Gamma_0$ & Variable $f$ and a flat prior on $\Gamma_0$&  [3.5,~~~6] &  [-35,\,-15]\\
 \ref{fig:T2_m_gamma_full_and_varfrac} (left) & f=1 $\Gamma_2$ & $f=100\%$ and flat prior on $\Gamma_2$&  [3.5,~~~6] &  [-30,\,0]\\
 \ref{fig:T2_m_gamma_full_and_varfrac} (right) & variable f $\Gamma_2$& Variable f and a flat prior on $\Gamma_2$ & [3.5,~~~6] & [-15,\,0]\\
 \bottomrule
\end{tabular}
\caption{
 \label{tab:priors}
List of priors on the CanDM parameters. All priors are logarithmic (i.e. flat priors on $\log_{10} m$).
The runs with $\alpha$ priors use a flat prior on $\log_{10}\alpha\in[1,4]$, the variable $f$ runs use a flat prior on $\log_{10} f \in [-2,0]$.
The $\Gamma_n$ and $\alpha$ prior lower limits were set where the model is essentially indistinguishable from $\Lambda$CDM, while the upper limits were just chosen sufficiently high to include all relevant parameter space (compare figures  \ref{fig:m_gamma_fullfrac} to \ref{fig:T2_m_gamma_full_and_varfrac}) 
Only the \enquote{tight} model $\Gamma_0$ priors were chosen narrowly to examine a specific region in parameter space.
The quantities $m$, $\Gamma_0$, and $\Gamma_2$ are taken in units of eV,  $\mathrm{eV}^{-7}$, and $\mathrm{eV}^{-5}$, respectively.
}
\end{table}

Our main result is shown in figure \ref{fig:f_a_varfrac}, where we show the joint constraint on the allowed interaction strength\footnote{We find that the bound on $\Gamma_0$ is (almost) mass-independent for any given fraction $f$ (see also figure \ref{fig:m_gamma_fullfrac}). As such, neither the $\Gamma_0$ vs $m$ or $m$ vs $f$ plots give any additional information, and we do not shown them.} parametrized through $\Gamma_0=\alpha^3/m^7$ and the fraction $f$. As expected, we find that the CMB data allows for increasingly large interaction strengths $\Gamma_0$ when the CanDM fraction $f$ decreases. However, higher fractions $f$ 
	can accommodate lower $S_8$ values while staying compatible with the CMB.
	This is because higher fractions 
	introduce a sharper and deeper step in the power spectrum (figure \ref{fig:Pk_suppression}), allowing a suppression at larger wavenumbers with less impact on the CMB to generate the same overall reduction of $S_8$. 
	
We consider next the case of a model with a temperature-dependent thermally averaged cross section, $\sigmavsq\propto T^2$ ($n=2$). Figure \ref{fig:T2_m_gamma_full_and_varfrac} shows the 
posterior distributions for this model (analogous to the right panel of figure \ref{fig:m_gamma_fullfrac} and figure \ref{fig:f_a_varfrac}).
The left panel shows the constraints on $m$ vs $\Gamma_2 = \alpha^3/m^5$ for $f=1$. As expected from equation
 \eqref{eq:Gammafactor}, $\Gamma_2$ is now the relevant parameter and the constraint is once again independent of $m$.
 The case of varying the fraction $f$ is shown in the right panel, where we can see that the results are similar to the $n=0$ case, with higher fractions allowing for lower $S_8$ values.
The comparison of the two cases highlights that the phenomenology of CanDM models barely depends on the precise theoretical setup.
 
 We found that in spite of its temporary contribution to $\Delta N_\mathrm{eff}$ around the time of recombination, CanDM does not help to solve the Hubble tension: our P18+lens+BAO results for $H_0$ are essentially identical in the $\Lambda$CDM and CanDM cases. At first sight, this is surprising, since the role of CanDM in the background evolution is qualitatively similar to that of early dark energy (EDE) \cite{Poulin:2018cxd,Agrawal:2019lmo}. However, solving the Hubble tension does not only require an increase in $N_\mathrm{eff}$ at early times, but also a special behavior at the level of perturbations, to avoid issues like additional Silk damping in the CMB spectrum tail. Some models involving EDE, self-interacting neutrinos \cite{Park:2019ibn}, or DM-DR interactions \cite{Buen-Abad:2017gxg} are able to match these requirements up to some extent, and to accommodate larger values of $H_0$. The CanDM model has a different behavior at the level of perturbations and does not appear to do so. We leave a more detailed investigation of this point and a comparison to other solutions of the Hubble tension for future work.

\subsection{Effect on the \texorpdfstring{$S_8$}{S8} tension}
\label{subsec:sigma8chi2}
We compare our results to the $S_8$ measurements from \cite{Joudaki:2019pmv}, which is a combination
of KiDS-VIKING-450 \cite{Wright:2018nix} and DES-Y1 \cite{Zuntz:2017pso, Drlica-Wagner:2017tkk} results
(KV-DES), yielding\footnote{More recent analyses with the KiDS-1000 + BOSS 3x2pt data
	\cite{Heymans:2020gsg}
	have been released during the final stages of this work. They find $S_8=0.766^{+0.20}_{-0.14}$ with a slightly higher central value but lower uncertainty and
	we would expect similar implications for CanDM.
} $S_8=0.762^{+0.025}_{-0.024}$.
Note that the $S_8$ value inferred from weak lensing measurements
generally depends on the assumed fiducial cosmology, however, here we neglect this subtlety and compare the $S_8$ values directly. To avoid correlations with other parameters,
we use $S_8=\sigma_8 \sqrt{\Omega_m/0.3}$ instead of $\sigma_8$.

\begin{figure}[t]
 \centering
 \includegraphics[width=\textwidth]{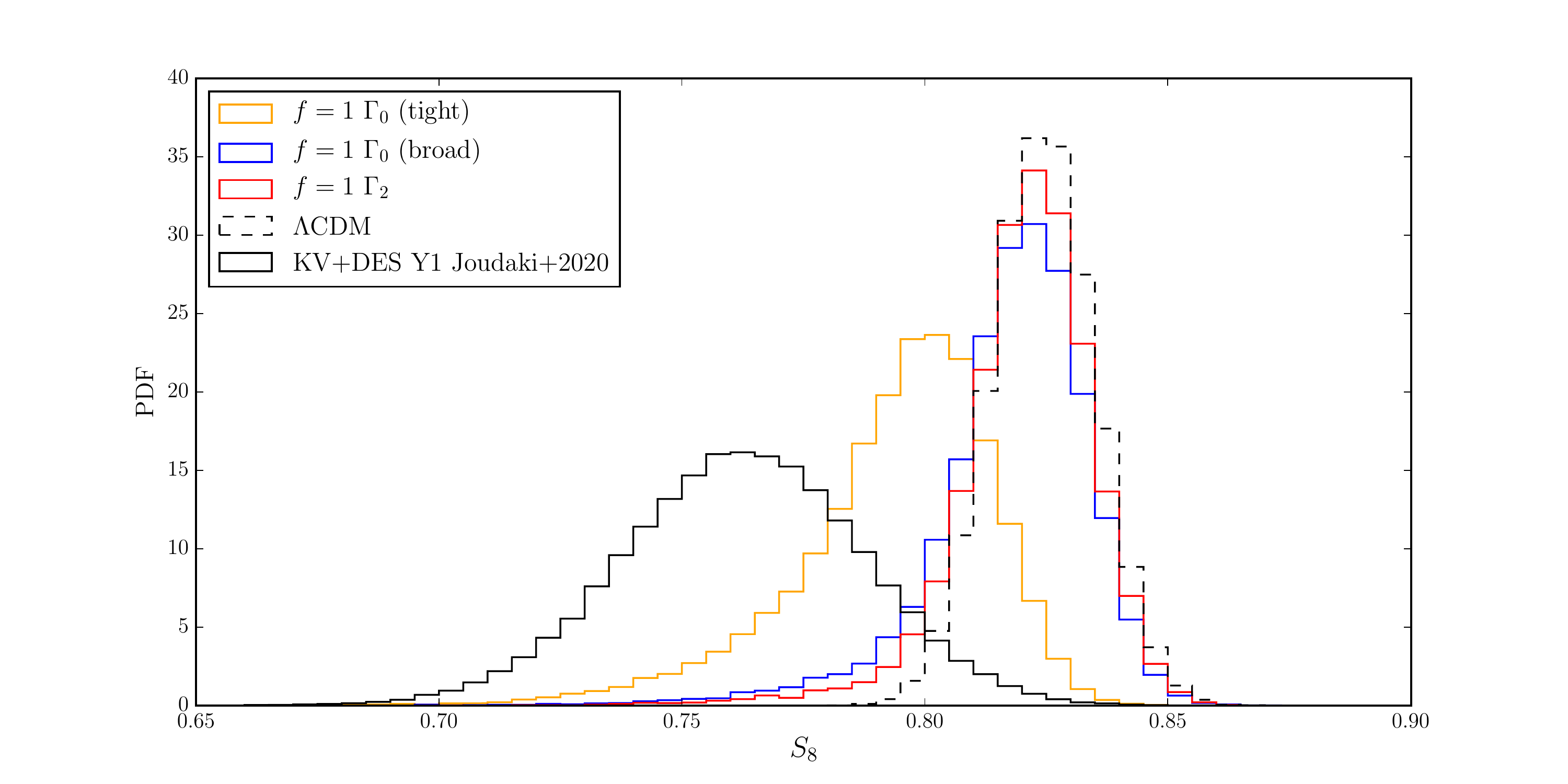}
 \caption{Histograms of the MCMCs showing the marginalized $S_8$ posteriors for five different combinations of models and data sets. We show three of the CanDM models of Table \ref{tab:priors}, and the $\Lambda$CDM  model in comparison. Additionally we display in black the Kids-Viking chains from Joudaki et al. 2020 \cite{Joudaki:2019pmv} labeled \enquote{KV+DES Y1 Joudaki+2020}.}
 \label{fig:chisquare}
\end{figure}

In figure \ref{fig:chisquare} we show the marginalized $S_8$ posterior for various prior choices in the CanDM model, compared with weak lensing measurements. In the small $\Gamma_n f^2$ limit, the CanDM model is equivalent to $\Lambda$CDM, and thus provides a very good fit to our P18+lens+BAO data. Then, with a top-hat prior on $\log \Gamma_n$ and possibly also on $\log f$, the fraction of the volume in parameter space in which $S_8$ is reduced (compared to the total volume favored by the data) depends crucially on the choice of lower prior edges. This can be checked by comparing the tight prior and broad prior cases in figure \ref{fig:chisquare}. Thus these posteriors do not fully account for the potential of the CanDM model to solve the $S_8$ tension.

However, we know from the previous section that CanDM models with $f \geq {\cal O}(10^{-1})$ are perfectly compatible with the low values of $S_8$ preferred by weak lensing data. It is thus legitimate to combine the P18+lens+BAO dataset with a weak lensing likelihood for such models. This will provide interesting statistical tests of the CanDM model, based on comparing the minimum $\chi^2$ for various models and combinations of  data.

\begin{figure}[t]
 \centering
 \includegraphics[width=0.6\textwidth]{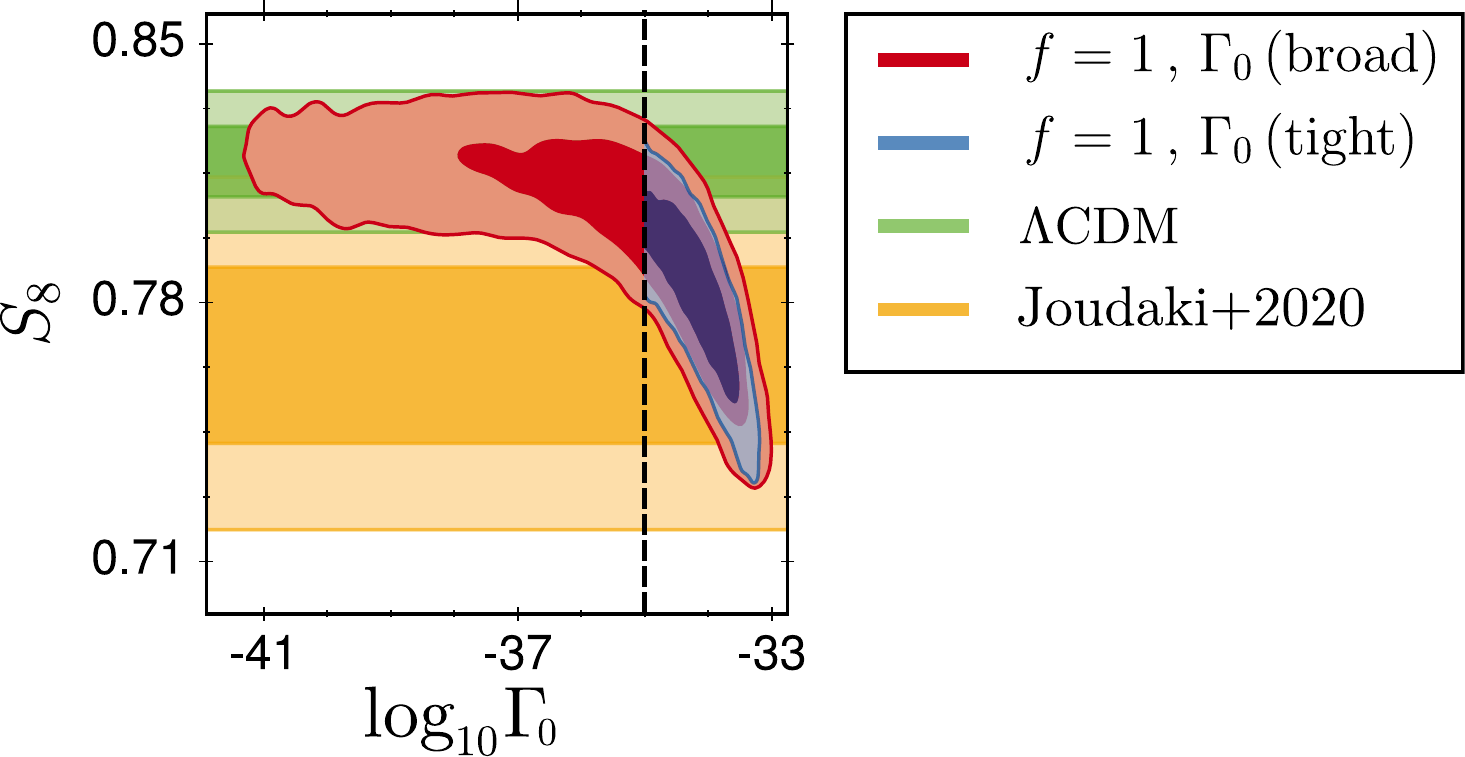}
 \caption{1 and 2$\sigma$ contours (68.3\% and 95.4\% CL) derived from  the P18+lens+BAO+WL chains.
  We show the case of CanDM with $f=1$  with broad prior in red, with tight prior in blue, and for $\Lambda$CDM in green. We additionally show the result from Joudaki+2020 \cite{Joudaki:2019pmv} as an orange band for comparison. The black dashed line shows the left prior edge of the tight prior on $\Gamma_0$. Note that the red contour is not only determined by $\Gamma_0>0$ being preferred by WL data: in this case part of the preference is also caused by our requirement that CanDM should have overlapping CE and NR phases (the gray wedge in figure \ref{fig:m_gamma_fullfrac}).}
 \label{fig:WL}
\end{figure}

We will assume that the analysis of weak lensing data by reference \cite{Joudaki:2019pmv} would give the same results for $S_8$ if CanDM was used as a baseline model instead of $\Lambda$CDM. We also assume that this measurement of $S_8$ is almost independent from that of other parameters. With such assumptions, we can model the results of \cite{Joudaki:2019pmv} as an independent likelihood, taking the form of an asymmetric gaussian probability distribution for $S_8=0.762^{+0.025}_{-0.024}$ (68\% confidence limit). We run new MCMC chains for the P18 dataset combined with this $S_8$ likelihood.
We show the marginalized posteriors in the two-dimensional $(S_8, \log_{10}\Gamma_0)$ parameter space in figure \ref{fig:WL}. The figure also shows how adopting a tighter prior on $\Gamma_0$ forces the chains into the region that allows to better solve the $S_8$ tension. 

In Table \ref{tab:chisq} we summarize the bestfit $\chi^2$ values of each model with each data combination. We derive from these numbers the level of tension between the P18+lens+BAO data and $S_8$ data in the framework of each CanDM model\footnote{The \tquote{internal} tension in this case is how much worse the combined fit becomes when both data sets are combined versus their individual fits, evaluating each fit within the given model. We use measures of $\Delta \chi^2$ for this as it naturally generalizes  the simplified $(\mu_1-\mu_2)/\sqrt{\sigma_1^2+\sigma_2^2}$ formula comparing two Gaussian distributions around means $\mu_i$ and with uncertainties $\sigma_i$ to completely arbitrary distributions.}. We also derive the global $\Delta \chi^2$ that shows to which extent each CanDM model performs better than $\Lambda$CDM.

The CanDM model can solve the  $S_8$ tension at the expense of a slightly worse fit to CMB+BAO. We see indeed that most CanDM models are able to absorb the $S_8$ tension (with $\chi^2_\mathrm{WL}<1$, instead of $\chi^2_\mathrm{WL}=5.1$ for $\Lambda$CDM), but only at the expense of increasing $\chi^2_\mathrm{PLB}$ by 3 to 5 units (where PLB is a short-cut notation for P18+lens+BAO)\footnote{In the model with the tightest prior, $\Delta \chi^2_\mathrm{PLB}$ is only of 0.6, but one starts from a $\chi^2_\mathrm{PLB}$ which is already worse than that of $\Lambda$CDM by 2.5 units, because in that case the \enquote{plain $\Lambda$CDM limit} is excluded by the choice of prior edges.}.
As a consequence, the best-fit $\chi^2$ for the combined dataset is lower with CanDM than with $\Lambda$CDM, but only by a small amount, ranging from -1.6 to -2.8. 
With two or three extra parameters in the CanDM models, this difference is very small. 
We conclude that the CanDM models are not significantly preferred by the data. 
Note that this conclusion depends on the fact that we choose a rather conservative measurement of $S_8$ by reference \cite{Joudaki:2019pmv}, only in 2$\sigma$ tension with Planck in the $\Lambda$CDM framework. 
Adopting lower or more constraining $S_8$ measurements -- such as \cite{Asgari:2019fkq} or \cite{Heymans:2020gsg} -- leading to a stronger tension would increase the preference for the CanDM models.
 
\begin{table}[h!]
 \centering
 \makebox[\textwidth][c]{
  \begin{tabular}{lcccc}
   \toprule
   Model & Min. $\chi_\mathrm{PLB}^2$ & $\Delta\chi_\mathrm{PLB+WL}^2$ & Tension & Global $\Delta\chi^2$ \\
   \midrule
   \lcdm                  & 2782.0 & 1.3 + 5.1 & 2.5 $\sigma$ &    0 \\
   $f=1$ $\alpha$           & 2782.4 & 1.7 + 2.0 & 1.9 $\sigma$ & -2.3 \\
   $f=1$ $\Gamma_0$ (broad) & 2782.8 & 3.4 + 0.3 & 1.9 $\sigma$ & -1.9 \\
   $f=1$ $\Gamma_0$ (tight) & 2784.5 & 0.6 + 0.5 & 1.0 $\sigma$ & -2.8 \\
   variable $f$ $\Gamma_0$  & 2782.5 & 3.3 + 0.5 & 1.9 $\sigma$ & -2.1 \\
   $f=1$ $\Gamma_2$         & 2781.4 & 4.6 + 0.8 & 2.3 $\sigma$ & -1.6 \\
   variable $f$ $\Gamma_2$  & 2781.7 & 3.8 + 1.2 & 2.2 $\sigma$ & -1.7 \\
   \bottomrule
  \end{tabular}
 }
 \caption{
  $\chi^2$ values and tensions for different configurations.
   The 2nd column shows the minimum $\chi^2$ of the P18+lens+BAO (PLB) likelihood for each model.
   The 3rd column shows the difference between the bestfit of the PLB+WL combination compared to that of PLB and WL only.
   The individual minimum of $\chi^2_{WL}$ is 0 by definition.
   The 4th column shows the remaining $S_8$ tension derived from the $\Delta\chi^2$ in the 3rd column.
   Finally the 5th column shows the \enquote{global} $\chi^2$ improvement of the model compared to $\Lambda$CDM, i.e.
   the minimum of $\chi_\mathrm{PLB+WL}^2$ in each model minus its value in row 1.
 \label{tab:chisq}
 }
\end{table}

\section{Conclusions}
\label{sec:conclusion}

In this paper we investigated the evolution of CanDM at the background and perturbation level, considering either a constant and or a temperature-dependent velocity-averaged cross section. We derived the corresponding evolution equations and implemented them within the Boltzmann solver \class. We demonstrated how CanDM can create a suppression in the matter power spectrum (similar to that induced by light warm dark matter or hot dark matter). This effect is caused by the reduced growth rate of dark matter density fluctuations below the CanDM Jeans length. We showed that this model is still compatible with CMB bounds, since it only impacts the CMB anisotropy spectra in a similar way as a small $\Delta N_\mathrm{eff}$.

We have computed strong exclusion limits on the interaction strength parameter $\Gamma_n=\alpha^3 m^{n-7} f^2$ for the two temperature scalings considered ($n=0$ and $n=2$),  as a function of the CanDM fraction $f$. We can summarize the constraints for $f=1$ as $\log_{10} \Gamma_0 \mathrm{eV}^{7}<-33.4$ and $\log_{10} \Gamma_2 \mathrm{eV}^{5}<-12.7$. 

We further studied the impact this model has on the $S_8$ tension, and we find that CanDM is a possible candidate to ease this tension, since models may have $S_8$ values as low as 0.76 or 0.77 and be still compatible with CMB+BAO data. This comes, however, at the expense of a slightly worse fit to CMB+BAO data, whose $\chi^2$ degrades by approximately 3 units. As long as the $S_8$ tension is moderate, solving it can at most reduce the weak lensing $\chi^2$ by 5 units. The two effects almost neutralize each other and the CanDM model (with its extra free parameters) does not appear as significantly preferred, and the combined $\chi^2$ only decreases by about 2-3 units. In the future, if the bounds on $S_8$ become tighter and in stronger tension with Planck (assuming the $\Lambda$CDM model), the CanDM model will remain as a possible explanation and will be preferred at a more significant level.

Lyman-$\alpha$ forest measurements such as \cite{Irsic:2017ixq,Palanque-Delabrouille:2019iyz} are an ideal probe of CanDM because they are sensitive to much smaller scales than the CMB. In particular, if CanDM were responsible for the suppression of structure at the
8 $\mathrm{Mpc}/h$ scale seen by weak lensing, it would lead to equal or stronger suppression on smaller scales.
We leave the study of Lyman-$\alpha$ bounds on CanDM parameters for future work, since it requires a modelling of the highly non-linear evolution of structures on very small scales. Furthermore, existing constraints such as \cite{deLaix+1995}\footnote{Note that this particular constraint would exclude the models with $f=1$, but could still ease the $S_8$ tension significantly for models with a fraction of around 10\%.} would require further investigation in the light of current and future Lyman-$\alpha$ data.

In summary, in this work, we presented the first full implementation of CanDM in a Boltzmann solver, and we studied its impact on different cosmological observables. While the model presents a rich phenomenology, mimicking both WDM and $\Delta N_\mathrm{eff}$ at different stages, current cosmological data show no statistically significant preference for this model over \lcdm. The model does not help solving the Hubble tension, but presents an interesting potential to solve the $S_8$ tension, if the latter gets confirmed by future measurements.
Future probes of the matter power spectrum on small scales will be highly sensitive to CanDM signatures and will either find stronger evidence for or rule out this model.

\section*{Acknowledgements}
This project was launched after illuminating discussions with Martin Schmaltz and Manuel Buen-Abad, who kindly provided the stand-alone code that they developed for their previous work on the same topic \cite{BuenAbad+2018}, as well as very useful insights and comments. 
SH acknowledges the support of STFC, via the award of a DTP Ph.D. studentship, and the Institute of Astronomy for a maintenance award.
NS acknowledges support from the DFG grant LE 3742/4-1. DH is supported by the FNRS research grant number~\mbox{F.4520.19}. JL is supported by the DFG grant LE 3742/3-1.
Simulations were performed with computing resources granted by RWTH Aachen University under project jara0184 and thes0567.

\appendix
\section{Thermodynamic equilibrium conservation laws}\label{ap:conservationLTE}

The homogeneous part of the Boltzmann equation implies 
\begin{equation}
	\parderconst{f}{t}{p} + \dot{p} \parderconst{f}{p}{t} = C[f]
\end{equation}
with collision operator $C[f]$. The moments of this equation can be calculated to find 
\begin{equation}
\label{eq:conservationequations}
\begin{alignedat}{1}
\dot{n} + 3Hn &= \dot{N}[f]~, \\
\dot{\rho} + 3H(\rho+P) &= \dot{Q}[f]~, \\
\dot{s} + 3 H s &= \dot{S}[f]~,
\end{alignedat}
\end{equation}
where we have used the standard definitions
\begin{alignat}{2}
n &= \int \frac{d^3p}{(2\pi)^3} f~, & 
\qquad \rho = \int \frac{d^3p}{(2\pi)^3} E f~, \qquad P = \int \frac{d^3p}{(2\pi)^3} \frac{p^2}{3E} f~, \label{eq:ap:nrhop}\\
s &= \int \frac{d^3p}{(2\pi)^3} s_B[f]~,  & \qquad s_B[f] = f \ln f \pm (1\mp f)\ln (1\mp f)~,
\end{alignat}
where the upper sign is for fermions and the lower sign for bosons. Correspondingly one finds the currents of the Boltzmann equation
\begin{alignat}{2}
\dot{N}[f] = \int \frac{d^3p}{(2\pi)^3}\,\, C[f]~, \qquad \dot{Q}[f] = \int \frac{d^3p}{(2\pi)^3} \, E \, C[f]~, \qquad \dot{S}[f] = \int \frac{d^3p}{(2\pi)^3} \ln\left(\frac{1\mp f}{f}\right)  C[f]~,
\end{alignat}
In local thermodynamic equilibrium (LTE) one can always find a Bose-Einstein or Fermi-Dirac distribution, both of which obey $\ln (\frac{1\mp f}{f}) = (E-\mu)/T$. Then one can immediately derive the relation
\begin{equation} \label{eq:entropyconservationcannibalsfull}
	\mathrm{LTE} \quad \Rightarrow \quad \dot{s}+3Hs = \frac{\dot{Q}[f]-\mu \dot{N}[f]}{T}~.
\end{equation}
From \fulleqref{eq:conservationequations} we can see that this is of course simply the first law of thermodynamics
\begin{equation}
	T \tfrac{d}{dt}\left(sa^3\right) = \dot{Q}[f]a^3 - \mu \dot{N}[f]a^3 = \tfrac{d}{dt}(\rho a^3) + P \tfrac{d}{dt}(a^3) - \mu \tfrac{d}{dt} (n a^3) \Leftrightarrow TdS = dU  + P dV - \mu dN~.
\end{equation}
Indeed, in LTE one can also show explicitly from the expression of $P$ in \fulleqref{eq:ap:nrhop} that 
\begin{equation}
	\tfrac{d}{dt} P = n \tfrac{d}{dt} \mu - s \tfrac{d}{dt} T \qquad  \Leftrightarrow \qquad S dT = - V dP + N d\mu~.
\end{equation}
Finally, this leads to the simple definition of entropy in LTE, namely
\begin{equation}
	s = \frac{\rho+P-\mu n}{T} \qquad \Leftrightarrow \qquad TS = U + VP - \mu N~.
\end{equation}

\begin{figure}[t]
	\centering
	\includegraphics[width=0.99\textwidth]{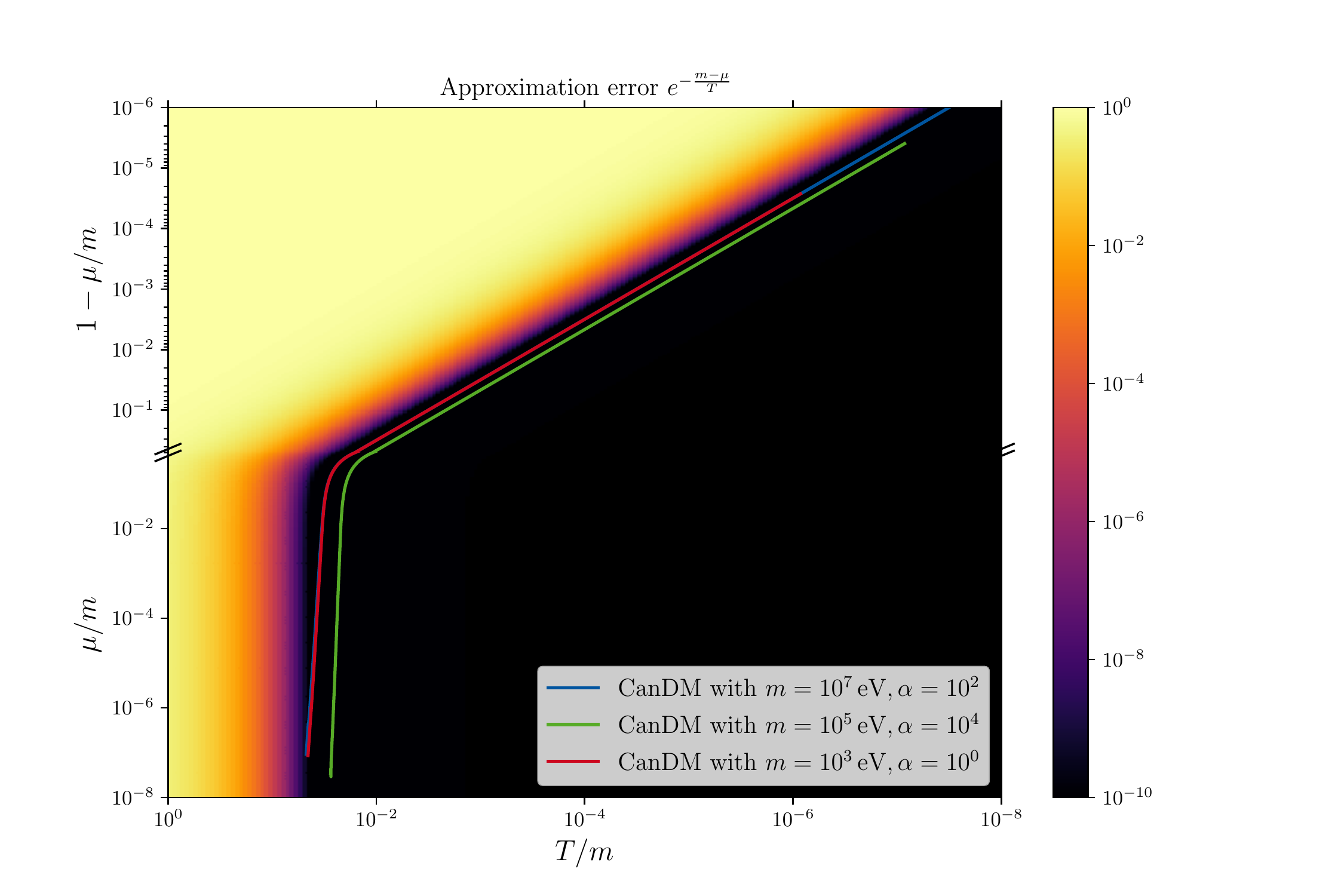}
	\caption{The figure shows where the Maxwell-Boltzmann distribution is a good approximation to the Bose-Einstein distribution. The color indicates the magnitude of the error made when computing $\rho$ or $P$ with the approximation \eqref{eq:bessel_approximation}. The $x/y$ axes show temperature/chemical potential in units of mass. Note that the upper part of the $y$ axis (for $\mu\approx M$) is plotted as $1-\mu/M$ to show the cold regime.
	The lines show three illustrative $\{T,\mu\}$-trajectories from where we use the NR approximation (bottom left) to where the evolution ends ($a=a_0$, top right).}
	\label{fig:Boltzmann_approximation}
\end{figure}

\subsection*{Application to CanDM}
For cannibalistic dark matter, one can easily find that $\dot{Q}[f] = 0$ from energy conservation during each reaction ($3\rightarrow2$ and $2\rightarrow2$), while $\dot{N}[f] \neq 0$ as long as the $3\rightarrow2$ reaction is still efficient. Then, using \fulleqref{eq:entropyconservationcannibalsfull} one finds that 
\begin{equation}
	\begin{alignedat}{1}
	&\dot{\rho} + 3H(\rho+P) = 0~,\\
	&\dot{n} + 3Hn = \dot{N}[f]~,\\
	&s = \frac{\rho+P-\mu n}{T} \quad \Rightarrow \quad \dot{s} + 3Hs = - \frac{\mu}{T} \dot{N}[f] = -\frac{\mu}{T} (\dot{n}+3Hn)~.
	\end{alignedat}
\end{equation}
In the case of CanDM, the number count current reads
\begin{equation} \label{eq:ap:dotN}
	\dot{N}[f] = \int \prod_{i=1}^{5} d\Pi_i \frac{\abs{M_{32}}^2}{12} (2\pi)^4 \delta^4\big({\textstyle\sum_j} p_j\big) \big(-f_1 f_2 f_3 + f_4 f_5 \big)~,
\end{equation}
where $d\Pi_i = \frac{d^3p_i}{2E_i}$, $M_{32}$ is the matrix element of the interaction from QFT, and the $f_i$ are the phase-space distributions of the interacting CanDM particles, with $f_i = 1/\left(\exp\left[(E_i-\mu)/T\right]\right)$ in LTE.

\section{Validity of the non-relativistic approximations}\label{ap:approximationvalidity}

\newcommand{\mmut}{\mathfrak{a}}
In section \ref{subsec:bgevo} we approximated the phase space distribution as a Maxwell-Boltzmann distribution to solve the CanDM evolution in the non-relativistic regime. This approximation relies mathematically on the fact that the exponent $(\sqrt{p^2+m^2}-\mu)/T$ is large. For this, it is sufficient that $T \ll m$ and the regions of non-relativistic evolution and chemical equilibrium overlap. We sketch out schematically the reason why this should be true below, and follow up this argument with a few numerical samples.

The exponent is bounded from below by $\mmut \equiv (m-\mu)/T$, and it is thus sufficient to show that $\mmut$ is large. In the cold phase of CanDM $\dot{N}[f] \to 0$, and the simultaneous conservation of entropy $s_\mathrm{NR}$ and number count $\nnr$ imply 
\begin{equation}
	const = \frac{s_\mathrm{NR}}{\nnr} = \frac{1}{T}\left[\frac{\rnr}{\nnr}+\frac{\pnr}{\nnr}-\mu\right] \approx 1+\frac{m-\mu}{T} = 1 + \mmut~,
\end{equation}
which gives an approximately constant $\mmut$ during this epoch. To find the size of this constant we can thus focus on the early non-relativistic regime where the efficient cannibalism drives $\mu \ll m$ and thus $\mmut \approx m/T$, which has $\mmut \gg 1$ by design. One could now suspect that in the intermediate region where neither $\mu \ll m$ nor $\dot{N}[f] \to 0$, one could violate $\mmut \gg 1$. In practice, there is not enough time for the $\mmut$ to deplete significantly.

In figure \ref{fig:Boltzmann_approximation} we show that the approximation has a negligible error for the
relevant $(T,\mu)$ evolution for a few selected models.


\bibliography{biblio}{}
\bibliographystyle{JHEP}

\end{document}